\documentclass[journal]{IEEEtran}
\usepackage{cite}
\pdfoutput=1
\ifCLASSINFOpdf
   \usepackage[pdftex]{graphicx}
   \graphicspath{{../figs/pdf/}{../figs/jpeg/}}
   \DeclareGraphicsExtensions{.pdf,.jpeg,.png}
\fi

\usepackage{amsmath}
\usepackage{amsfonts}
\usepackage{xcolor}
\usepackage{booktabs}

\usepackage{multirow}
\ifCLASSOPTIONcompsoc
 \usepackage[caption=false,font=normalsize,labelfont=sf,textfont=sf]{subfig}
\else
 \usepackage[caption=false,font=footnotesize]{subfig}
\fi

\usepackage{url}
\usepackage{xspace}
\usepackage{amsmath}

\DeclareMathOperator*{\argmin}{arg\,min}
\begin{document}
\bstctlcite{IEEEexample:BSTcontrol}
% @IEEEtranBSTCTL{IEEEexample:BSTcontrol,
% CTLdash_repeated_names = "no",
% CTLuse_forced_etal       = "yes",
% CTLmax_names_forced_etal = "4",
% CTLnames_show_etal       = "1" }

%
% paper title
% Titles are generally capitalized except for words such as a, an, and, as,
% at, but, by, for, in, nor, of, on, or, the, to and up, which are usually
% not capitalized unless they are the first or last word of the title.
% Linebreaks \\ can be used within to get better formatting as desired.
% Do not put math or special symbols in the title.
\title{Missing MRI Pulse Sequence Synthesis using Multi-Modal Generative Adversarial Network}
\author{Anmol~Sharma,~\IEEEmembership{Student~Member,~IEEE,}
        Ghassan~Hamarneh,~\IEEEmembership{Senior~Member,~IEEE}% <-this % stops a space
\thanks{This work was partially supported by the NSERC-CREATE Bioinformatics 2018-2019 Scholarship.}%
\thanks{Anmol Sharma and Ghassan Hamarneh are with the Medical Image Analysis Laboratory, School of Computing Science, Simon Fraser University, Canada. 
e-mail: \{asa224, hamarneh\}@sfu.ca}% <-this % stops a space
}%
% \thanks{Corresponding author: Anmol Sharma}

% The paper headers
\ifCLASSOPTIONpeerreview
    \markboth{Transaction on Medical Imaging}%
    {IEEE TMI}
\else
    % \markboth{IEEE Journal}%
    % {Second Header Target Journal}
    \markboth{}%
    {}
\fi

\newcommand{\tone}{$T_1$\xspace} 
\newcommand{\tonec}{$T_{1c}$\xspace} 
\newcommand{\ttwo}{$T_2$\xspace} 
\newcommand{\ttwof}{$T_{2flair}$\xspace}
\newcommand{\mmgan}{MM-GAN\xspace}
\newcommand{\allseq}{\tone, \tonec, \ttwo, \ttwof}
\newcommand{\G}{$\mathcal{G}$\xspace}
\newcommand{\D}{$\mathcal{D}$\xspace}
\newcommand{\pptone}{$\text{P2P}_{T_1}$\xspace}
\newcommand{\ppttwo}{$\text{P2P}_{T_2}$\xspace}
\newcommand{\pp}{$\text{P2P}$\xspace}

\newcommand{\pgtone}{$\text{pGAN}_{T_1}$\xspace}
\newcommand{\pgttwo}{$\text{pGAN}_{T_2}$\xspace}
\newcommand{\pg}{$\text{pGAN}$\xspace}

\newcommand{\mitone}{$\text{MI-GAN}_{T_1}$\xspace}
\newcommand{\mittwo}{$\text{MI-GAN}_{T_2}$\xspace}
\newcommand{\mi}{$\text{MI-GAN}$\xspace}
\newcommand{\red}[1]{\textcolor{black}{#1}}
\newcommand{\redtwo}[1]{\textcolor{black}{#1}}

% If you want to put a publisher's ID mark on the page you can do it like
% this:
%\IEEEpubid{0000--0000/00\$00.00~\copyright~2015 IEEE}
% Remember, if you use this you must call \IEEEpubidadjcol in the second
% column for its text to clear the IEEEpubid mark.
% use for special paper notices
%\IEEEspecialpapernotice{(Invited Paper)}

% make the title area
% \newcommand{\red}[1]{\textcolor{violet}{#1}}
\maketitle
% %.......
% \IEEEpubid{\begin{minipage}{\textwidth}\ \\[45pt] \centering
% This work has been submitted to the IEEE for possible publication. Copyright may be transferred without notice, after which this version may no longer be accessible.
% \end{minipage}} 
% %.......
\begin{abstract}
Magnetic resonance imaging (MRI) is being increasingly utilized to assess, diagnose, and plan treatment for a variety of diseases. The ability to visualize tissue in varied contrasts in the form of MR pulse sequences in a single scan provides valuable insights to physicians, as well as enabling automated systems performing downstream analysis. However many issues like prohibitive scan time, image corruption, different acquisition protocols, or allergies to certain contrast materials may hinder the process of acquiring multiple sequences for a patient. This poses challenges to both physicians and automated systems since complementary information provided by the missing sequences is lost. In this paper, we propose a variant of generative adversarial network (GAN) capable of leveraging redundant information contained within multiple available sequences in order to generate one or more missing sequences for a patient scan. The proposed network is designed as a multi-input, multi-output network which combines information from all the available pulse sequences and synthesizes the missing ones in a single forward pass. We demonstrate and validate our method on two brain MRI datasets each with four sequences, and show the applicability of the proposed method in simultaneously synthesizing all missing sequences in any possible scenario where either one, two, or three of the four sequences may be missing. We compare our approach with competing unimodal and multi-modal methods, and show that we outperform both quantitatively and qualitatively.
\end{abstract}

% and the consequences may range from hindered decision making in manual analysis to outright algorithm failure in automated systems due to their strong implicit dependence on a set of available pulse sequence

% Note that keywords are not normally used for peerreview papers.
\begin{IEEEkeywords}
generative adversarial networks, multi-modal, missing modality, pulse sequences, MRI, synthesis.
\end{IEEEkeywords}

% For peer review papers, you can put extra information on the cover
% page as needed:
% \ifCLASSOPTIONpeerreview
% \begin{center} \bfseries EDICS Category: 3-BBND \end{center}
% \fi
%
% For peerreview papers, this IEEEtran command inserts a page break and
% creates the second title. It will be ignored for other modes.
\IEEEpeerreviewmaketitle

\section{Introduction}
% form to use if the first word consists of a single letter:
% \IEEEPARstart{A}{demo} file is ....

\IEEEPARstart{M}{edical} imaging forms the backbone of the modern healthcare systems, providing means to assess, diagnose, and plan treatments for a variety of diseases. Imaging techniques like computed tomography (CT), magnetic resonance imaging (MRI), X-Rays have been in use for over many decades. Magnetic resonance imaging (MRI) out of these is particularly interesting in the sense that a single MRI scan is a grouping of multiple pulse sequences, each of which provides varying tissue contrast views and spatial resolutions, without the use of radiation. These sequences are acquired by varying the spin echo and repetition times during scanning, and are widely used to show  pathological changes in internal organs and muscoskeletal system. Some of the commonly acquired sequences are \tone-weighted, \ttwo-weighted, \tone-with-contrast-enhanced (\tonec), and \ttwo-fluid-attenuated inversion recovery (\ttwof), though there exist many more \cite{Jackson97_Pulse_Sequences}.

A combination of sequences provide both redundant and complimentary information to the physician about the imaged tissue, and certain diagnosis are best performed when a particular sequence is observed. For example, \tone and \ttwof sequences provide clear delineations of the edema region of tumor in case of glioblastoma, \tonec provides clear demarcation of enhancing region around the tumor used as an indicator to assess growth/shrinkage, and \ttwof sequence is used to detect white matter hyperintensities for diagnosing vascular dementia (VD) \cite{Bowles2016}. 

In clinical settings, however, it is common to have MRI scans acquired using varying protocols, and hence varying sets of sequences per patient. Sequences which are routinely acquired may be unusable or missing altogether due to scan corruption, artifacts, incorrect machine settings, allergies to certain contrast agents and limited available scan time~\cite{Havaei2016_HeMIS, Varsavsky2018_PIMSS, Chartsias2018}. This phenomenon is problematic for many downstream data analysis pipelines that assume presence of a certain set of pulse sequences to perform their task. For instance, most of the segmentation methods \cite{Chen2016, MICCAI2017_proceedings, Isensee2017, Wang2018, Dolz2018} proposed for brain MRI scans depend implicitly on the availability of a certain set of sequences in their input in order to perform the task. Most of these methods are not designed to handle missing inputs, and hence may fail in the event where some or most of the sequences may be absent.

Modifying existing pipelines in order to handle missing sequences is hard, and may lead to performance degradation. Also, the option of redoing a scan to acquire the missing/corrupted sequence is impractical due to the expensive nature of the acquisition, longer wait times for patients with non-life-threatening cases, need for registration between old and new scans, and rapid changes in anatomy of area in-between scan times due to highly active abnormalities such as glioblastoma. Hence there is a clear advantage in retrieving any missing sequence or an estimate thereof, without having to redo the scan or changing the downstream pipelines. 

To this end, we propose a multi-modal generative adversarial network (\mmgan) which is capable of synthesizing missing sequences by combining information from all available sequences. The proposed method exhibits the ability to synthesize, with high accuracy, all the required sequences which are deemed missing in a single forward pass through the network. The term ``multi-modal" simply refers to the fact that the GAN can take multiple-modalities of available information as input, which in this case represents different pulse sequences. Similar to the input being multi-modal, our method generates multi-modal output containing synthesized versions of the missing sequences. Since most of the downstream analysis pipelines commonly target $C=4$ pulse sequences $S$ = \{\tone, \tonec, \ttwo, $T_{2flair}$\} as their input \cite{Pereira2016, Havaei2017, MICCAI2017_proceedings}, we design our method around the same number of sequences, although we note that our method can be generalized to any number $C$ and set $S$ of sequences. The input to our network is a 4-channel (corresponding to $C=4$ sequences) 2D axial slice, where a \red{zero image} is imputed for channels corresponding to missing sequences. The output of the network is a 4-channel 2D axial slice, in which the originally missing sequences are synthesized by the network. 

The rest of the paper is organized as follows: Section \ref{sec:relatedWork} presents a review of the MR sequence synthesis literature. Section \ref{sec:contributions} provides an overview of the key contributions of this work. Section \ref{sec:methodology} presents the proposed method in detail. Section \ref{sec:experimentalSetup} provides details about the method implementation, datasets used, as well as outlines experimental setup for the current work. Section \ref{sec:discussion} discusses the results and observations for the proposed method, and finally the paper is concluded in Section \ref{sec:conclusion}. 

\section{Related Work}
\label{sec:relatedWork}
% first tell the reader that there have a lot of interest in MR sequence synthesis, citing all the papers. 
There has been an increased amount of interest in developing methods for synthesizing MR pulse sequences \cite{Iglesias2013synthesizing, Ye2013, Jog2013, Jog2014_RF_FLAIR, Van2015does, Jog2015tree, Roy2016, Yawen2016, Bowles2016, Sevetlidis2016, Jog2017_REPLICA, Chartsias2018, Olut2018_GAN, Mehta2018_RS, Salem2019_UNet, Yu2019_EAGan}. We present a brief overview of previous work in this field by covering them in two sections: Unimodal, where both the input and output of the system is a single pulse sequence \red{(one-to-one)}; and multimodal, where methods are able to leverage multiple input sequences to synthesize a single \red{(many-to-one)} or multiple sequences \red{(many-to-many)}. 

\subsection{Unimodal Synthesis}
In unimodal synthesis \red{(one-to-one)}, a common strategy includes building an atlas or a database that maps intensity values between given sequences. Jog et al.~\cite{Jog2013} used a bagged ensemble of regression trees trained from an atlas. The training data $(\mathcal{A}_1, \mathcal{A}_2)$ consisted of multiple image patches $\mathcal{A}_1$ around a voxel $i$ in a source sequence, and a single intensity value at the same voxel in a target sequence, as $\mathcal{A}_2$. The use of image patches to predict the intensity value of a single voxel in output sequence allows representing many-to-many relationship between intensity values of input and target sequences. Ye et al.~\cite{Ye2013} propose an inverse method, which performs a local patch-based search in a database for every voxel in the target pulse sequence. Once the patches are found, they are ``fused" together using a data-driven regularization approach. Another atlas based method was proposed in~\cite{Roy2016} where \ttwo whole-head sequence (including skull, eyes etc.) is synthesized from the available \tone images. The synthesized \ttwo sequence is used to correct distortion in diffusion-weighted MR images by using it as template for registration, in the absence of a real \ttwo sequence. Yawen et al.~\cite{Yawen2016} leverage joint dictionary learning (JDL) for synthesizing any unavailable MRI sequence from available MRI data. JDL is performed by minimizing the inconsistency between statistical distributions of the dictionary codes for input MRI sequences while preserving the geometrical structure of the input image. 

Supervised machine learning and deep learning (DL) based methods have also been employed in sequence synthesis pipelines. A 3D continuous-valued conditional random field (CRF) is proposed in~\cite{Jog2015tree} to synthesize \ttwo images from \tone. The synthesis step is encoded as a maximum a-posterior (MAP) estimate of Gaussian distribution parameters built from a learnt regression tree. Nguyen et al.~\cite{Nguyen2015} was one of the first to employ DL in the form of location-sensitive deep network (LSDN) for sequence synthesis. LSDN predicts the intensity value of the target voxel by using voxel-centered patches extracted from an input sequence. The network models the responses of hidden nodes as a product of feature and spatial responses. Similarly, Bowleset et al.~\cite{Bowles2016} generate ``pseudo-healthy" images by performing voxel-wise kernel regression instead of deep networks to learn local relationships between intensities in \tone and \ttwof sequences of healthy subjects. Since most of the methods were based on local features in the form of patches and did not leverage global features of the input sequence, Sevetlidis et al.~\cite{Sevetlidis2016} proposed an encoder-decoder style deep neural network trained layer-wise using restricted Boltzmann machine (RBM) based training. The method utilized global context of the input sequence by taking a full slice as input. Recently, Jog et al.~\cite{Jog2017_REPLICA} propose a random forest based method that learns intensity mapping between input patches centered around a voxel extracted from a single pulse sequence, and the intensity of corresponding voxel in target sequence. The method utilized multi-resolution patches by building a Gaussian pyramid of the input sequence. Yu et al.~\cite{Yu2019_EAGan} propose a unimodal GAN architecture to synthesize missing pulse sequences in a one-to-one setting. The approach uses an edge detection module that tries to preserve the high-frequency edge features of the input sequence, in the synthesized sequence. Recently, Ul Hassan Dar et al.~\cite{Salman2019_TMI} propose to use a conditional GAN to synthesize missing MR pulse sequences in a unimodal setting for two sequences \tone and \ttwo. 

\subsection{Multimodal Synthesis}
Multimodal synthesis has been a relatively new and unexplored avenue in MR synthesis literature. One of the first multi-input, single-output \red{(many-to-one)} method was proposed by Jog et al.~\cite{Jog2014_RF_FLAIR}; a regression based approach to reconstruct \ttwof sequence using combined information from \tone, \ttwo, and proton density (PD) sequences. Reconstruction is performed by a bagged ensemble of regression trees predicting the \ttwof voxel intensities. Chartsias et al.~\cite{Chartsias2018} were one of the first to propose a multi-input, multi-output \red{(many-to-many)} encoder-decoder based architecture to perform many-to-many sequence synthesis, although their \red{multimodal} method is tested only using a single-output \red{(\ttwof) (many-to-one setting)}. Their network is trained using a combination of three loss functions, and uses a feature fusion step in the middle that separates the encoders and decoders. Olut et al.~\cite{Olut2018_GAN} present a GAN based framework to generate magnetic resonance angiography (MRA) sequence from available \tone, and \ttwo sequences. The method uses a novel loss function formulation, which preserves and reproduces vascularities in the generated images. Although for a different application, Mehta et al.~\cite{Mehta2018_RS} proposed a multi-task, multi-input, multi-output 3D CNN that outputs a segmentation mask of the tumor, as well as a synthesized version of \ttwof sequence. The main aim remains to predict tumor segmentation mask from three available sequences \tone, \ttwo, and \tonec, and no quantitative results for \ttwof synthesis using \tone, \ttwo, and \tonec are provided.

Though all the methods discussed above propose a multi-input method, none of the methods have been proposed to synthesize multiple missing sequences (multi-output), and in one single pass. All three methods~\cite{Jog2014_RF_FLAIR},~\cite{Chartsias2018}, and~\cite{Mehta2018_RS} synthesize only one sequence \red{(either \ttwof or \ttwo, many-to-one setting)} in the presence of varying number of input sequences, while \cite{Olut2018_GAN} only synthesizes MRA using information from multiple inputs \red{(many-to-one)}. Although the work presented in ~\cite{Olut2018_GAN} is close to our proposed method, theirs is not a truly multimodal network (many-to-many), since there is no empirical evidence that their method will generalize to multiple scenarios. \red{Similarly, the framework proposed in~\cite{Chartsias2018} can theoretically work in a many-to-many setting, but no empirical results are given to demonstrate its scalability and applicability in a variety of different scenarios, as we do in this work. The authors briefly touch upon this by adding a new decoder to already trained many-to-one network, but do not explore it any further}. To the best of our knowledge, we are the first to propose a method that is capable of synthesizing multiple missing sequences using a combination of various input sequences \red{(many-to-many)}, and demonstrate the method on the complete set of scenarios (i.e., all combinations of missing sequences).

The main motivation for most synthesis methods is to retain the ability to meaningfully use some downstream analysis pipelines like segmentation or classification despite the partially missing input. However, there have been efforts by researchers working on those analysis pipelines to bypass any synthesis step by making the analysis methods themselves robust to missing sequences. Most notably, Havaei et al.~\cite{Havaei2016_HeMIS} and Varsavsky et al.~\cite{Varsavsky2018_PIMSS} provide methods for tumor segmentation using brain MRI that are robust to missing sequences~\cite{Havaei2016_HeMIS}, or to missing sequence labels~\cite{Varsavsky2018_PIMSS}. Although the methods bypass the requirement of having a synthesis step before actual downstream analysis, the performance of these robust versions of analysis pipelines often do not match the state-of-the-art performance of other non-robust methods in the case when all sequences are present. This is due to the fact that the methods not only have to learn how to perform the task (segmentation/classification) well, but also to handle any missing input data. This two-fold objective for a single network raises a trade-off between robustness and performance. 

\section{Contributions}
\label{sec:contributions}
The following are the key contributions of this work:
\begin{enumerate}
    \item We propose the first \redtwo{empirically validated} \red{multi-input multi-output} MR pulse sequence synthesizer capable of synthesizing missing pulse sequences using \emph{any combination of available sequences as input} without the need for tuning or retraining of models, \red{in a many-to-many setting}.
    \item \red{The proposed method is capable of synthesizing \emph{any combination of target missing sequences as output} in one single forward pass, and requires only a single trained model for synthesis. This provides significant savings in terms of computational overhead during training time compared to training multiple models in the case of unimodal and multi-input single-output methods.}
    \item \red{We propose to use implicit conditioning (IC), a combination of three design choices, namely imputation in place of missing sequences for input to generator, sequence-selective loss computation in the generator, and sequence-selective discrimination. We show that IC improves overall quantitative synthesis performance of generator compared to the baseline approach without IC.}
    % \item We demonstrate that the proposed method does not need explicit supervision or conditioning in order to synthesize a particular missing sequence. Instead, the proposed method relies on a combination of three design choices for implicit conditioning, namely imputed noise in the input in place of missing sequences, sequence-selective loss computation in the generator, and sequence-selective discrimination. 
    \item To the best of our knowledge, we are the first to incorporate curriculum learning based training for GAN by varying the difficulty of examples shown to the network during training. 
    \item Through experiments, we show that we outperform both the current state-of-art in unimodal (REPLICA~\cite{Jog2017_REPLICA} and pGAN~\cite{Salman2019_TMI}), as well as the multi-input single-output synthesis (MM-Synthesis~\cite{Chartsias2018}) method. We also set up new benchmarks on a complete set of scenarios using the BraTS2018 dataset. 
\end{enumerate}

\begin{figure*}[t]
    \centering
    \includegraphics[scale=0.47]{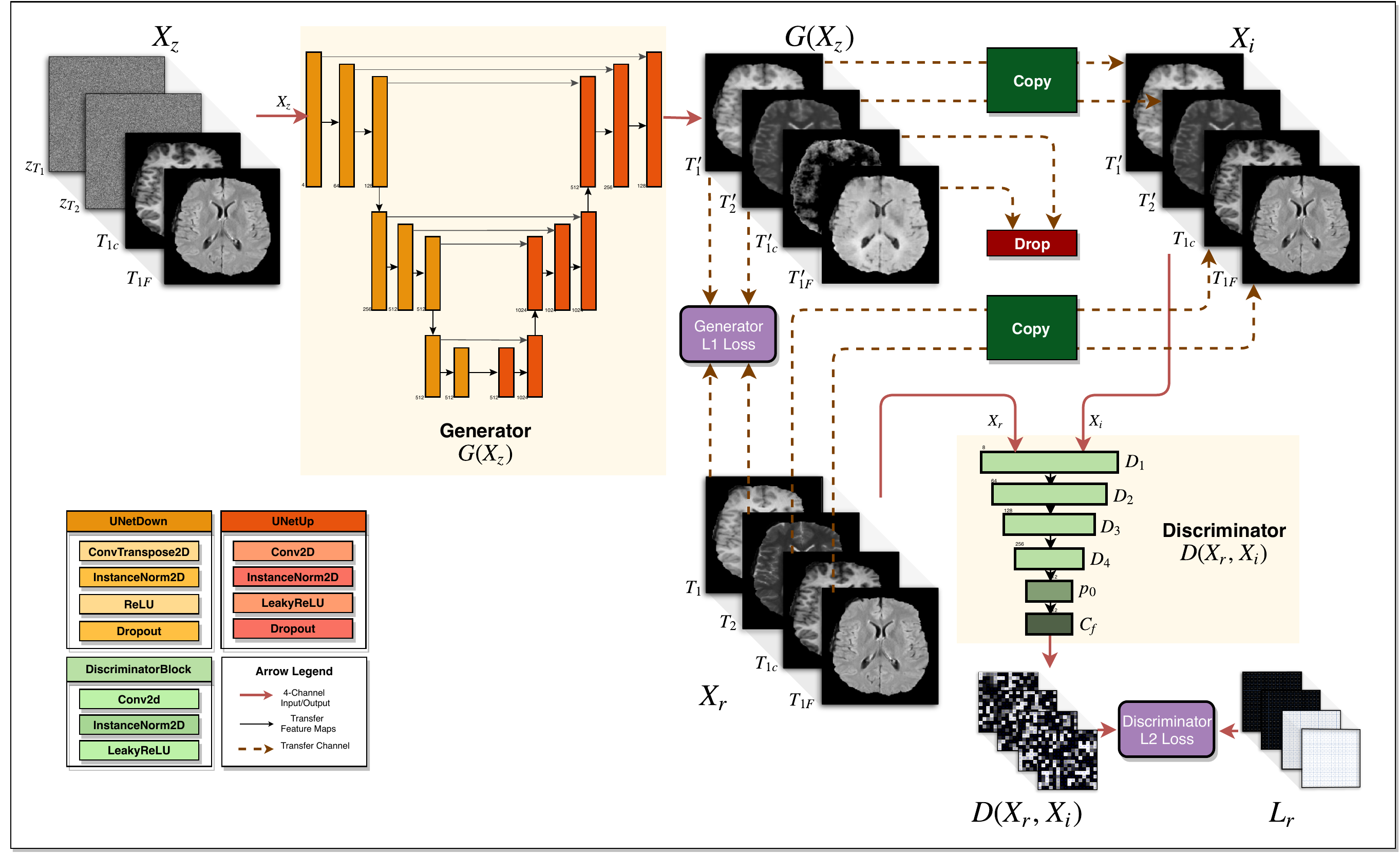}
    \caption{Proposed multimodal generative adversarial network (MM-GAN) training process. \red{The generator is a UNet architecture, while the discriminator is a PatchGAN architecture with L2 loss (least squares GAN). The green ``Copy" blocks transfer the input channels as is to its output, while the red ``Drop" block deletes its input channels. The generator L1 loss is computed only between the synthesized versions of the missing sequences (here \tone and \ttwo). The discriminator takes $X_r$ and $X_i$ as input and produces a 4-channel 2D output representing whether each $N\times N$ patch in $X_i$ is either fake or real.}}
    \label{fig:proposed}
\end{figure*}

\section{Methodology}
\label{sec:methodology}
\subsection{Background}
Generative adversarial networks (GANs) were first proposed by Goodfellow et al. \cite{Goodfellow_GAN2014} in order to generate realistic looking images. 
A GAN is typically built using a combination of two networks: generator (\G) and discriminator (\D).
The generator network is tasked with generating realistic data, typically by learning a mapping from a random vector $z$ to an image $I$, $\mathcal{G}: z \rightarrow I$,  where $I$ is said to belong to the generator's distribution $p_{\mathcal{G}}$. 
The discriminator 
$\mathcal{D}: I \rightarrow t$ maps its input $I$ to a target label $t\in\{0,1\}$, where $t=0$ if $I \in p_{\mathcal{G}}$, i.e. a fake image generated by $\mathcal{G}$ and $t=1$ if $I \in p_{r}$ where $p_r$ is the distribution of real images. 
A variant of GANs, called conditional-GAN (cGAN) \cite{Mirza2014_cGAN}, proposes a generator that learns a mapping from a random vector $z$ and a class label $y$ to an output image $I \in p_{\mathcal{G}}$, $\mathcal{G}: (z, y) \rightarrow I$. Another variant of cGAN called Pix2Pix~\cite{Isola2017_Pix2Pix} develops a GAN in which the generator learns a mapping from an input image $x \in p_r$ to output image $I \in p_{\mathcal{G}}$, $\mathcal{G}: x \rightarrow I$, and the discriminator learns a mapping from two input images, $x_1$ and $x_2$, to $T$, $\mathcal{D}: (x_1, x_2) \rightarrow T$. $x_1$ and $x_2$ may belong to either $p_r$ (real) or $p_{\mathcal{G}}$ (fake). The output $T$ in this case is a not a single class label, but a binary prediction tensor representing whether each $N\times N$ patch in the input image is real or fake~\cite{Isola2017_Pix2Pix}.

A GAN is trained in an adversarial setting, where the generator (parameterized by $\theta_{\mathcal{G}}$) is trained to synthesize realistic output that can ``fool" the discriminator into classifying them as real, and the discriminator (parameterized by $\theta_{\mathcal{D}}$) is trained to accurately distinguish between real data and fake data synthesized by the generator. GAN input/outputs can be images~\cite{Isola2017_Pix2Pix}, text~\cite{Reed2016_GAN_Text} or even music~\cite{Yang_2017_MiDiNet}. 
Both the generator and discriminator act as adversaries to each other, hence the training formulation forces both networks to continuously get better at their tasks.
GANs found tremendous success in a variety of different tasks, ranging from face-image synthesis \cite{Gauthier2014_project}, image stylization~\cite{Ulyanov2016_instance}, future frame prediction in videos \cite{Mathieu2015_deep}, text-to-image synthesis~\cite{Reed2016_GAN_Text} and synthesizing scene images using scene attributes and semantic layout~\cite{Karacan2016_Learning}. 
GANs have also been utilized in medical image analysis ~\cite{Kazeminia2018_Survey}, particularly for image segmentation ~\cite{Moeskops2017_GAN, Xue2018_SegAN, Izadi2018_Skin}, normalization~\cite{Bentaieb2018_Hist}, synthesis~\cite{Olut2018_GAN, Yu2019_EAGan, Salman2019_TMI} as well as image registration ~\cite{Tanner2018_Generative}. 

\subsection{Proposed Method}
\label{sub:proposedmethod}
We propose a variant of Pix2Pix architecture~\cite{Isola2017_Pix2Pix} called Multi-Modal Generative Adversarial Network (MM-GAN) for the task of synthesizing missing MR pulse sequences in a single forward pass while leveraging all available sequences. The following subsections would outline the detailed architecture of our model.

% \begin{figure*}[t]
%     \centering
%     \includegraphics[scale=0.45]{figs/pdf/proposed_test.pdf}
%     \caption{Proposed multimodal generative adversarial network (MM-GAN) training process. }
%     \label{fig:proposed}
% \end{figure*}

\subsubsection{Generator}
\label{sec:G}
The generator of the proposed method is a UNet~\cite{Ronneberger2015_UNET}, which has proven useful in a variety of segmentation and synthesis tasks due to its contracting and expanding paths in the form of encoder and decoder subnetworks. 
 The architecture is illustrated in Figure~\ref{fig:proposed}.
 The convolution kernel sizes for each layer in the generator is set to $4\times4$.
 The generator network is a combination of \texttt{UNetUp} and \texttt{UNetDown} blocks.
 The input to the generator is a 2D axial slice from a patient scan with $C=4$ channels representing four pulse sequences, and spatial size of $256\times256$ pixels.
 The network is designed with a fixed input size of 4-channels, where channel $C=0,1,2,\text{and } 3$ corresponds to \tone, \ttwo, \tonec, and \ttwof, respectively. \redtwo{Hence for any $C$-sequence 3D scan, the proposed method works sequentially on $C$-channel 2D axial slices.}
 In order to synthesize missing sequences, the channels corresponding to each missing sequence are imputed with \red{zeros}.
 The imputed version (along with the real sequences) becomes the input to the generator and is represented by $X_z$.
 For instance, if sequences \tone and  \ttwo are missing, channels $C=0$ and $C=1$ in the input image are imputed with a \red{zero} image of size $256\times256$.
 The output of the generator is given by $\mathcal{G}(X_z | \theta_{\mathcal{G}})$ and is of the same size as the input.
 Due to design, the generator always outputs 4 channels, however, as we outline in the subsequent text the output channels corresponding to the existing real sequences are not used for loss computation and are replaced with the real sequences before relaying them as input to the discriminator.
 During training the ground truth image $X_r$,
 short for ``real'', which is of the same size as $X_z$ contains all ground truth sequences at their respective channel indices. We use the term ``image" for a single 2D slice with 4 channels.
 
MM-GAN observes both an input image and \red{imputed zeros $z$} in the form of $X_z$, in contrast to vanilla Pix2Pix where the generator is conditioned just by an observed image $x$. The reasons behind this design choice are discussed in subsection \ref{subsubsec:implicit}. \red{We also investigate different imputation strategies in Suppl. Mat. Section I-A, and found that zero based imputation performs the best quantitatively.}
 
To optimize $\theta_{\mathcal{G}}$, our generator adopts the general form of the generator loss in Pix2Pix, which is a combination of a reconstruction loss $\mathcal{L}_1$ and an adversarial loss $\mathcal{L}_2$ used to train the generator to fool the discriminator, i.e.

\begin{equation}
    \begin{split}
        \theta_{\mathcal{G}}^{*} = \argmin_\mathcal{\theta_{\mathcal{G}}}\:
        \lambda &\mathcal{L}_{1}(\mathcal{G}(X_z | \theta_{\mathcal{G}}), X_r)+\\
        (1 - \lambda)\:&\mathcal{L}_{2}(\mathcal{D}(X_i, X_r | \theta_{\mathcal{D}}), L_{ar}).\\
    \end{split}
    \label{eq:generatorLoss}
\end{equation}

To calculate $\mathcal{L}_1$, we select synthesized sequences from $\mathcal{G}(X_z | \theta_{\mathcal{G}})$, that were originally missing, and compute the L1 norm of the difference between the synthesized versions of the sequence and the available ground truth from $X_r$. Mathematically, given the set $K$ containing the indices of missing sequences (e.g. $K=\{0, 2\}$ when \tone and \tonec are missing) in the current input, we calculate $\mathcal{L}_{1}$ only for the sequences that are missing ($k=0, 2$), and sum the values. 

% \begin{equation}
%     \mathcal{L}_{1}(X_r, \mathcal{G}(X_z | \theta_{\mathcal{G}})) = \sum_{k\in K} \left(  |X_r^k - \mathcal{G}(X_z | \theta_{\mathcal{G}})^k|  \right).
% \end{equation}

To calculate $\mathcal{L}_2$, we compute the squared L2 norm of the difference between the discriminator's predictions $\mathcal{D}(X_i, X_r | \theta_{\mathcal{D}}))$
and a dummy ground truth tensor $L_{ar}$ of the same size as the output of \D. In order to encourage the generator to synthesize sequences that confuse or ``fool" the discriminator into predicting they are real, we set all entries of $L_{ar}$ to ones, masquerading all generated sequences as real. $X_i$ is introduced in the next section.

The choice of L1 as a reconstruction loss term for the generator is motivated by its ability to prevent too much blurring in the final synthesized sequences, as compared to using an L2 loss (similar to~\cite{Isola2017_Pix2Pix}). 

\subsubsection{Discriminator}
\label{sec:D}
% write about architecture of discriminator
We use the PatchGAN architecture~\cite{Isola2017_Pix2Pix} for the discriminator part of our MM-GAN. PatchGAN architecture learns to take into account the local characteristics of its input, by predicting a real/fake class for every $N\times N$ patch of its input, compared to classic GANs where the discriminator outputs a single real/fake prediction for the whole input image. This encourages the generator to synthesize images not just with proper global features (shape, size), but also with accurate local features (texture, distribution, high-frequency details).  
 In our case we set $N=16$.
%  This promotes the generator to learn to generate high-frequency (sharp) details of the synthesized images \cite{Isola2017_Pix2Pix} in order to successfully fool the discriminator, since the discriminator works at patch scale.
 
The discriminator is built using four blocks followed by a zero padding layer and a final convolutional layer (Figure~\ref{fig:proposed}).
 The convolutional kernel sizes, stride and padding is identical to the values used in the generator (subsection~\ref{sec:D}).
 Due to the possibility of having a varying number of sequences missing, instead of providing just the synthesized sequences and their real counterparts as input to the discriminator, we first create a modified version of $\mathcal{G}(X_z|\theta_{\mathcal{G}})$ by dropping the \red{reconstructed} sequences that were originally present, and replacing them with the original sequences from $X_r$.
 The modified version of $\mathcal{G}(X_z|\theta_{\mathcal{G}})$ is represented by $X_i$, short for ``imputed".
 The input to the discriminator is a concatenation of $X_i$ and $X_r$.
 This is also illustrated in Figure \ref{fig:proposed}.
 
 The discriminator is trained to output a 2D patch of size $16\times16$ pixels, with 4 channels corresponding to each sequence.
 In order to supervise the discriminator during training, we use a 4-channel 2D image based target, in which each channel corresponds to a sequence. 
 More specifically, given missing sequences $K$ (e.g., $K=\{0,2\}$, \tone and \tonec missing), the target (i.e. ground truth) variable for $\mathcal D$ is
 $L_r^k=\{0.0\;\text{(fake)}\;\text{if}\;k \in K,\;\text{else}\;1.0\;\text{(real)}\}$.
  Note that $L_r^k$ is a 2D tensor of size $16\times16$ (since each 256$\times$256 image is divided into 16$\times$16 patches) yet the assignment of 0.0 or 1.0 represents an assignment to the whole $16\times16$ $L_r^k$ tensor (since the whole image is either real or fake and not patch-specific). This is also illustrated in Figure \ref{fig:proposed}.
  
 Between the output of discriminator $\mathcal{D}(X_i, X_r|\theta_{\mathcal{D}})$ and $L_r$, an L2 loss is computed.
 The final discriminator loss becomes:
 \begin{equation}
     \begin{split}
         \theta_{\mathcal{D}}^{*} = \argmin_{\theta_{\mathcal{D}}}\;& \mathcal{L}_{2}(\mathcal{D}(X_r, X_r|\theta_{\mathcal{D}}), L_{ar})\\ + & \mathcal{L}_{2}(\mathcal{D}(X_r, X_i|\theta_{\mathcal{D}}), L_r).
     \end{split}
 \end{equation}
\red{This is equivalent to a least-squares GAN since the loss function incorporates an L2 loss. }
\subsubsection{Implicit conditioning}
\label{subsubsec:implicit}
\red{Due to the inherent design of deep learning architectures, the input as well as output of a convolutional neural network model has to have a fixed channel dimension. In our use case however, both the input and output channel dimensions vary (since the number of available sequences can vary). In order to address this problem, we propose a combination of three design choices, which we collectively refer to as implicit conditioning (IC). In IC, the varying input channels problem is solved by imputing a fixed value (zeros) to the input channels where the sequences are missing. For the problem of generator output size being fixed in channel dimension, one possible approach can be to synthesize all four input sequences. The loss function can be calculated between four ground truth sequences, and the four synthesized sequences. However, this poses a challenge for the generator, as its burdened with generating all sequences, including the reconstruction  of  the  ones  that  were  provided as input. In order to address this, we proposed selective loss computation in \G, where the loss is only calculated between the ground truth sequences that were missing, and the corresponding output channels of the generators. In conjunction, we also propose selective discrimination in \D, which ensured stability during training by preventing the discriminator from overpowering the generator. We also show that IC-based training outperforms the baseline training methodology of generating and penalizing inaccurately synthesizing all sequences (Suppl. Mat. Section I-B). The design choices are individually summarized below.}

% In order to make the generator generalize to all possible scenarios where one or more sequences may be missing, it is imperative that the generator has the ability to recognize which sequence is actually missing, and which ones are present.
%  Previously, approaches like the cGAN \cite{Mirza2014_cGAN} explicitly condition the generator as a ``supervision" to generate an image of a particular class.
%  However, in this work, we note that this may not be necessary, and the conditioning can be performed implicitly using a combination of three design choices: \red{input} imputation in place of missing sequences for input to \G; selective loss computation in \G; and selective discrimination in \D. \red{Moreover, the major contribution of implicit conditioning lies in relieving the generator from the burden of generating all sequences including the reconstruction of the ones that were provided as input. The selective loss computation part caters to this problem, and allows the network to concentrate only on successfully synthesize the missing sequences, as it is not penalized for inaccurately reconstructing the input sequences.}\\
 \noindent \textsc{\red{Input} imputation}: The input $X_z$ of the generator always \red{contains an imputed value ($z=$ zeros)} in place of the missing sequences which acts as a way to condition the generator and informs which sequence(s) to synthesize.\\
 \noindent \textsc{Selective loss computation in \G}: In conjunction, the $\mathcal{L}_{1}(\mathcal{G})$ loss that is computed only between the synthesized sequences for the generator, and then backpropagated, allows the generator to align itself towards only synthesizing the actual missing sequences properly while ignoring the performance in synthesizing the ones that were already present.\\
 \noindent \textsc{Selective discrimination in \D}: Imputing real sequences at the generator output (i.e. $X_i$) before providing it as discriminator input forces the discriminator to accurately learn to delineate only between the synthesized sequences and their real counterparts.
 Since the generator loss function also has a term that tries to fool the discriminator, this allows selective backpropagation into the generator where it is penalized only for incorrectly synthesizing the missing sequences, and not for incorrectly synthesizing the sequences that were already present.
 This relieves the generator of the difficult task of synthesizing all sequences in the presence of some sequences.
 
%  The major contribution of implicit conditioning lies in relieving the generator from the burden of generating all sequences including the reconstruction of the ones that were provided as input. The selective loss computation part caters to this problem, and allows the network to concentrate only on successfully synthesize the missing sequences, as it is not penalized for inaccurately reconstructing the input sequences.
 \subsubsection{Curriculum learning}
\label{subsubsec:cl}
In order to train our proposed method we use a curriculum learning (CL) \cite{Bengio2009_Curriculum} based approach.
\red{In CL based training, the network is initially shown easier examples followed by increasingly difficult examples as the training progresses. We hypothesized that CL can benefit in training of MM-GAN due to an ordering in the level of difficulty across the various scenarios that the network has to handle. If some cases are ``easier" than others, it might be useful if the easier cases are shown first to the network in order to allow the network to effectively learn when ample supervision is available. As the network trains, ``harder" cases can be introduced so that the network adapts without diverging. In our work, scenarios with 1 sequence missing are considered ``easy", followed by a ``moderate" set of scenarios with 2 missing sequences, and lastly, the scenarios with 3 missing sequences are considered ``hard". We adopted this ordering in our work, and showed the network easier examples first, followed by moderate and finally hard examples. After a threshold of 30 epochs, we show every scenario with uniform probability. }

%  In CL based training, the network is initially shown easier examples followed by increasingly difficult examples as the training progresses.
%  In our case, the difficulty of examples increases as more sequences go missing. Initially, the network is shown scenarios with only one missing sequence. As the training progresses, scenarios with two or more sequences missing are sampled for training. After a certain threshold epoch as the training nears the end, all scenarios are made equally likely to be shown to the network during training. 

%  Since we primarily deal with four sequences in all our experiments, we assign a 4-bit string for each scenario, where the index of each bit represents a particular sequence.
%  The order of bits is preserved in the sense that the first bit always represents the first sequence in the input to generator.
%  A value of 0 at a bit index means that the sequence is missing, and is to be synthesized by the generator.
%  A value of 1 indicates a sequence that is present.
%  The total number of scenarios for a 4-bit representation is 16, however two scenarios (0000, and 1111) are invalid in our case.
%  Hence the network is trained with 14 scenarios, where easier scenarios are shown at the commencement of the training, followed by harder scenarios as the training progresses.
%  It is to be noted that the bit representation is only for concise formulation and visualization purposes, and in no way used during training/testing phases of the method.
 
\section{Experimental Setup}
\label{sec:experimentalSetup}
In this section we describe different aspects of the experiments that are performed in this work. 

\subsection{Datasets}
In order to validate our method we use brain MRI datasets from two sources, namely the Ischemic Stroke Lesion Segmentation Challenge 2015 (ISLES2015)\red{~\cite{Maier2017_ISLES}} and the Multimodal Brain Tumor Segmentation Challenge 2018 (BraTS2018)\red{~\cite{Menze2015}}.\\[-4mm]

\textit{1) ISLES2015} dataset is a publicly available database with multi-spectral MR images~\cite{Maier2017_ISLES}.
 We choose the sub-acute ischemic stroke lesion segmentation (SISS) cohort of patients, which contains 28 training and 36 testing cases.
 The patient scans are skull stripped \red{using BET2~\cite{Jenkinson2005_BET2}}, and resampled to an isotropic resolution of $1$~mm$^3$.
 Each scan consists of four sequences namely \tone, \ttwo, DWI, and \ttwof, and are \red{rigidly} co-registered to the \ttwof sequence \red{using elastix tool-box~\cite{Klein2010_Elastix}}. More information about the preprocessing steps can be found in the original publication~\cite{Maier2017_ISLES}.
 We use 22 patients from the SISS training set for experiments. \\[-4mm]
 
\textit{2) BraTS2018} consists of a total of 285 patient MR scans acquired from 19 different institutions,  divided into two cohorts: \red{glioblastoma/high grade glioma (GBM/HGG) and low grade glioma (LGG)}.
 The patient scans contains four pulse sequence \tone, \ttwo, \tonec, and \ttwof.
 All scans are resampled to $1$~mm$^3$ isotropic resolution \red{using a linear interpolator}, skull stripped, and co-registered with a single anatomical template \red{using rigid registration model with mutual information similarity metric}. \red{Detailed preprocessing information can be found in~\cite{Menze2015}}. In order to demonstrate our method's ability in synthesizing sequences with both high grade and low grade glioma tumors present, we use a total of 210 patients from HGG and 75 \red{patients} from LGG cohort. 195 patients are reserved for training for HGG cohort, while 65 are used for training in LGG experiments. For validation, we use 5 patients for both HGG and LGG cohorts. In order to test our trained models, we use 10 patients from HGG cohort (due to larger data available), while we report results using 5 patients for LGG cohort as testing. 
  
\subsection{Preprocessing}
Each patient scan is normalized by dividing each sequence by its mean intensity value. This ensures that distribution of intensity values is preserved~\cite{Chartsias2018}. \red{Normalization by mean is less sensitive to high or low intensity outliers as compared to min-max normalization procedures, which can be greatly exacerbated by the presence of just a single high or low intensity voxel in a sequence. This is especially common in the presence of various pathologies, like tumors as in BraTS2018 datasets, which tend to have very high intensity values in some sequences (\ttwo, \ttwof) and recessed intensities in others (\tone, \tonec). In practice, we observed this for BraTS2018 HGG cohort, where some voxels had an unusually high intensity value due to a pathology. On performing min-max normalization to scale intensities between [0,1], we found that the presence of very high intensity voxel squashed the pixel range to always lie very close to zero. This artificially bumped the performance numbers for the generator since most voxels lied close to zero, and hence the generator could synthesize images with intensity values close to zero, and achieve a low L1 score easily. On the other hand, mean normalization was relatively unaffected due to a large number of voxels in a defined range (~0-4000). The mean value was not strongly affected by the presence of one or more high/low intensity voxels. 
We also tested the method internally with zero mean and unit variance based standardization, and found the results to be at par with mean normalization.}
 In order to crop out the brain region from each sequence, we calculate the  largest bounding box that can accommodate each brain in the whole dataset, and then use the coordinates to crop each sequence in every patient scan.
 The final cropped size of a single patient scan with all sequences contains 148 axial slices of size $194\times155$.
 Each slice in every sequence is resized to a spatial resolution of $256\times256$, using bilinear interpolation, in order to maintain compatibility with UNet architecture of the generator. 
 \red{We note that avoiding resampling twice (once during registration performed by the original authors of the dataset, and once during resampling to $256\times 256$ in this work) may preserve some information in the scans that may otherwise be lost. However, it is a necessary preprocessing step in order to maintain compatibility with various network architectures that we utilize, which includes inherent assumptions that the input size would be a power of two to allow successive contracting and expanding steps used in many encoder-decoder style architectures. In order to fully avoid the second resampling step (to $256\times 256$ in this work), a different network architecture may be used without the encoder-decoder setup, though the performance of those networks may not be at par with the modern encoder-decoder style networks as established in synthesis field~\cite{Salman2019_TMI, Isola2017_Pix2Pix, Johnson2016_Perceptual}}

\subsection{Benchmark Methods}
We compare our method with three competing methods, one unimodal and two multimodal.
 The unimodal (single-input, single-output\red{, one-to-one}) method we compare against is pGAN \cite{Salman2019_TMI}, while the multimodal \red{(many-to-one)} models being REPLICA \cite{Jog2017_REPLICA} (in a multi-input setting), and that of  Chartsias et al.~\cite{Chartsias2018}, called MM-Synthesis hereafter. \red{Both pGAN and MM-Synthesis were recently published (2019 and 2018), and they outperform all other methods before them (MM-Synthesis outperforms LSDN~\cite{Nguyen2015}, Modality Propagation~\cite{Ye2013}, and REPLICA~\cite{Jog2017_REPLICA}, while pGAN outperforms both REPLICA and MM-Synthesis in one-to-one synthesis). To the best of our knowledge, we did not find any other methods that claimed to outperform either pGAN or MM-Synthesis, and so decided to evaluate our method against them.  
}
 For comparison with pGAN \cite{Salman2019_TMI}, we reimplement the method using \red{the open source code provided with the publication}, and train both pGAN and our method on a randomly chosen subset of data from BRATS2018 LGG cohort. \red{We also compare with a standard baseline which is a vanilla Pix2Pix~\cite{Isola2017_Pix2Pix} model trained and tested in a one-to-one setting}. For our multimodal \red{(many-to-one)} experiments, \red{we report mean squared error (MSE) results for both REPLICA and MM-Synthesis directly from \cite{Chartsias2018}, as we recreate the exact same testbed for comparison with \mmgan, as used in MM-Synthesis. We adopt the} same testing strategy (5-fold cross validation), database (ISLES2015), and scenarios (7 scenarios where \ttwof is always missing and is the only one that is synthesized). \red{As highlighted in \cite{Chartsias2018}, the multi-input version of REPLICA required seven models each for each of the seven valid scenarios in many-to-one setting synthesizing \ttwof sequence. MM-Synthesis and our proposed \mmgan only required a single multimodal (many-to-one) network which generalized to all seven scenarios}. 
 For our final extended set of experiments, \red{we demonstrate the effectiveness of our method in a multi-input multi-output (many-to-many) setting, where} we perform testing on the HGG and LGG cohorts of BRaTS2018 dataset for which we report results of all 14 valid scenarios (=16$-$2, as scenario when all sequences are missing/present are invalid for our experiments) instead of just 7. \red{The results showcase our method's generalizability on different use-cases with varying input and output subsets of sequences, and different difficulty levels. We use a fixed order of sequences (\tone, \ttwo, \tonec, \ttwof) throughout this paper, and represent each scenario as a 4-bit string, where a zero (0) represents the absence of a sequence at that location, while a one (1) represents its presence}. 
 
\subsection{Training and Implementation Details}
In order to optimize our networks, we use Adam~\cite{Kingma2014_ADAM} optimizer with learning rate $\eta=0.0002$, $\beta_1=0.5$ and $\beta_2=0.999$. Both the generator and discriminator networks are initialized with weights sampled from a Gaussian distribution with $\mu=0, \sigma=0.02$.

\red{We perform four experiments, first for establishing that multi-input synthesis is better than single-input (many-to-one vs one-to-one respectively), second for \ttwof synthesis using multiple inputs (many-to-one) (called MISO, short for multi-input single-output) using ISLES2015 dataset to compare with REPLICA and MM-Synthesis. The third set of experiments encompasses validation of multiple key components proposed throughout this paper, in terms of their contribution towards overall network performance. We test different imputation strategies ($z=\{average, noise, zeros\}$), as well as the effect of curriculum learning (CL) and implicit conditioning (IC). These are included in the supplementary materials accompanying this manuscript, in Suppl. Mat. Section I. The final set of experiments pertain to multimodal synthesis (MIMO, short for multi-input multi-output), which sets a new benchmark for many-to-many synthesis models using BraTS2018 HGG and LGG cohorts. We refer to the second and fourth experiments as MISO and MIMO, respectively, hereafter.}
 We use a batch size of 4 slices to train models, except for MISO, where we use batch size of 2.
 We train the models for 30 epochs in MISO and 60 epochs for MIMO sets of experiments, with no data augmentation.
 Both the generator and discriminator networks are initialized with weights sampled from Gaussian distribution with $\mu=0, \sigma=0.02$.
 
 We choose $\lambda=0.9$ for the generator loss given in equation \ref{eq:generatorLoss}, while we multiply the discriminator loss by 0.5 which essentially slows down the rate at which discriminator learns compared to generator.
 During each epoch, we alternate between a single gradient descent step on the generator, and one single step on the discriminator.
 
 For our MIMO experiments, we use the original PatchGAN~\cite{Isola2017_Pix2Pix} discriminator.
 However for our MISO experiments, due to lack of training data, we used a smaller version of the PatchGAN discriminator with just two discriminator blocks, followed by a zero padding and final convolution layer.
 Also, random noise was added to both $X_r$ and $X_i$ inputs of the discriminator in MISO experiments. 
 This was done to reduce the complexity of the discriminator to prevent it from overpowering the generator, which we observed when original PatchGAN with no noise imputation in its inputs was used for this set of experiments.
 The generator's final activation was set to ReLU for MIMO and linear for MISO experiments due to the latter having negative intensity values for some patients.
 
 For our implementation we used Python as our main programming language.
 We implemented Pix2Pix architecture in PyTorch.
 The computing hardware consisted of an i7 CPU with 64 GB RAM and GTX1080Ti 12 GB VRAM GPU.
 Throughout our experiments we use random seeding in order to ensure reproducibility of our experiments.
 For our MIMO experiments, we use curriculum learning by raising the difficulty of scenarios every 10 epochs (starting from one missing sequence) that are shown to the network until epoch 30 (shown examples with three missing sequences), after which the scenarios are shown randomly with uniform probability until epoch 60.
 For MISO experiments we train the model without curriculum learning, and show all scenarios with uniform probability to the network for 30 epochs. MM-GAN takes an average time of 0.1536 $\pm$ 0.0070 seconds per patient as it works in constant time at test-time w.r.t number of sequences missing.

\subsection{Evaluation Metrics}
Evaluating the quality of synthesized images should ideally take into account both the quantitative aspect (per pixel synthesis error) as well as qualitative differences mimicking human perception. In order to cover this spectrum, we report results using three metrics, namely mean squared error (MSE), peak signal-to-noise ratio (PSNR) and structural similarity index metric (SSIM). The MSE is given as $\frac{1}{n}\sum_{i=1}^{N}(y_i - y_{i}^{'})^2$ where $y_i$ is the original sequence and $y_{i}^{'}$ is the synthesized version. MSE however depends heavily on the scale of intensities, hence for fair comparison, similar intensity normalization procedures were followed. In this work, we adopt the normalization procedure used in~\cite{Chartsias2018}. \red{We report all results except in Section \ref{subsec:multimodal} after normalizing both ground truth and synthesized image in range [0, 1]. We do this in order to maintain consistency across the study, and allow all future methods to easily compare with our reported values regardless of the standardization/normalization procedure used in network training. We note that the generator successfully learns to synthesize images that lie in the same normalized range as the ground truth and input training images, and hence there is no need for re-normalization after synthesis. Re-normalization in our case was only applied before evaluation to ensure fair comparison for current and future works. 
For Section \ref{subsec:multimodal}, in order to directly compare with the results reported in~\cite{Chartsias2018}, we report results without re-normalizing}. In order to still provide a normalization agnostic metric, we report PSNR, which takes into account both the MSE and the largest possible intensity value of the image, given as: $10\log_{10}\left(I_{max}^{2}/\text{MSE}\right)$, where $I_{max}$ is the maximum intensity value that the image supports, which depends on the datatype. We also report SSIM, which tries to capture the human perceived quality of images by comparing two images. SSIM is given as: $\frac{(2\mu_x\mu_y + c_1)(2\sigma_{xy} + c_2)}{(\mu_x^2 + \mu_y^2 + c_1)(\sigma_x^2 + \sigma_y^2 + c_2)}$, where $x,y$ are two images to be compared, and $\mu=$mean intensity, $\sigma^2=$variance of image, and $\sigma_{xy}=$covariance of $x,y$. 

\section{Results and Discussion}
\label{sec:discussion}
\red{In this section we present the results for our experiments validating our method and comparison with competing unimodal (one-to-one) (Section \ref{subsec:unimodal}), and multi-input single-output (many-to-one) methods (Section \ref{subsec:multimodal}) methods. 
Finally we present benchmark results in multi-input multi-output (MIMO) synthesis in Section \ref{subsec:mimo}.}

\subsection{Single-Input VS Multi-Input Synthesis}
\label{subsec:unimodal}
In order to understand the performance difference between using a single sequence versus using information from multiple sequences to synthesize a missing sequence, we set up an experiment evaluating the two approaches. Our hypothesis for this experiment is that multiple sequences provide complimentary information about the imaged tissue, and hence should be objectively better than just using one sequence for the task of synthesizing a missing one. 
We set up an experiment to compare \red{multi-input single-output model with two single-input single-output models in two tasks, namely synthesizing missing \tone and \ttwo sequences respectively. For single-input single-output models, we set up a Pix2Pix~\cite{Isola2017_Pix2Pix} model as baseline, called P2P. We also compare with a state-of-art single-input single-output synthesis method called pGAN~\cite{Salman2019_TMI}. pGAN adopts it's generator architecture from~\cite{Johnson2016_Perceptual} and proposes a combination of L1 reconstruction loss and perceptual loss using VGG16 as the loss-network in a generative adversarial network framework. The discriminator for pGAN was adopted from~\cite{Isola2017_Pix2Pix}. We use the official pGAN implementation\footnote{https://github.com/icon-lab/pGAN-cGAN} for training and testing on the pGAN (k=3) model. Finally, for our multi-input single-input model, we implement a multi-input single-output variant of our proposed MM-GAN called MI-GAN (multi-input GAN).}

\red{We call the baseline Pix2Pix models $\text{P2P}_{T_1}$ (synthesizing \tone from \ttwo) and $\text{P2P}_{T_2}$ (synthesizing \ttwo from \tone). Similarly, pGAN models are named as $\text{pGAN}_{T_1}$ and $\text{pGAN}_{T_2}$. For our multi-input variants, the two variants of MI-GAN are named as $\text{MI-GAN}_{T_1}$ and $\text{MI-GAN}_{T_2}$, which synthesize \tone from (\ttwo, \tonec, \ttwof) and \ttwo from (\tone, \tonec, \ttwof) respectively. For P2P and MI-GAN models training was performed for 60 epochs, using consistent set of network hyperparameters used throughout this paper. For pGAN, training was performed as outlined in the original paper \cite{Salman2019_TMI} for 100 epochs, with k=3 slices as input to the network. All networks were trained on 70 patients from LGG cohort of BraTS2018 dataset, and tested on 5 patients. Although input normalization between the P2P/MI-GAN and pGAN differ, all metrics were calculated after normalizing both ground truth and synthesized image in range [0, 1]. We also perform Wilcoxon signed-rank tests across all test patients and report p-values wherever the performance difference is statistically significant ($p<$ 0.05). }

 % -------------------------------------------------------------------------------
% -------------------------------------------------------------------------------
\begin{figure}[t]
    \centering
    \includegraphics[width=0.35\textwidth]{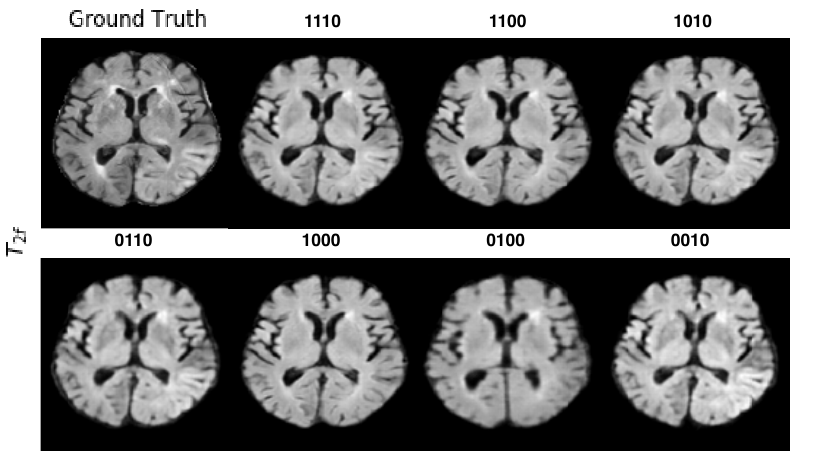}
    \caption{Qualitative results from \ttwof synthesis experiments with ISLES2015 dataset. The order of scenario bit-string is \tone, \ttwo, DW, \ttwof\red{, where a zero (0) represents missing sequence, while a one (1) represents an available sequence. We note that the presence of multiple sequences (scenario 1110) qualitatively matches the closest to the ground truth, with sharp boundaries between white and grey matter, as well as less blurring artifacts compared to scenarios with two or just one available sequences.} Patient images shown here are from VSD ID \emph{70668} from ISLES2015 SISS cohort.}
    \label{fig:islesqual}
\end{figure}
% -------------------------------------------------------------------------------
% -------------------------------------------------------------------------------

 % -------------------------------------------------------------------------------
% -------------------------------------------------------------------------------
\begin{figure*}[t]
    \centering
    \includegraphics[width=0.78\textwidth]{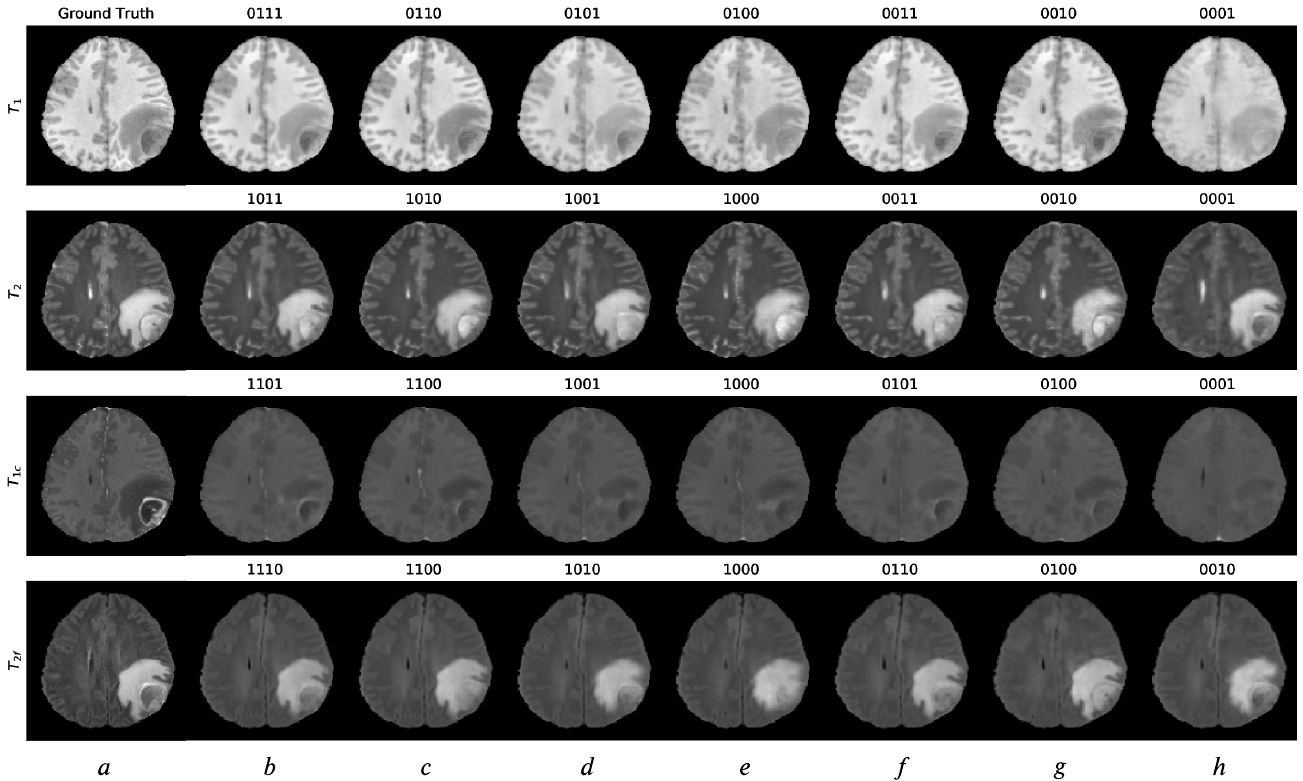}
    \caption{Qualitative results from multimodal synthesis experiments with BRaTS2018 dataset. Each row corresponds to a particular sequence \red{(row names on the left in order \tone, \ttwo, \tonec and \ttwof). Columns are indexed at the bottom of the figure by alphabets (a) through (h), and have a column name written on top of each slice. Column names are 4-bit strings where a zero (0) represents missing sequence that was synthesized, and one (1) represents presence of sequence. Column (a) of each row shows the ground truth slice, and the subsequent columns ((b) through (h)) show synthesized versions of that slice in different scenarios. The order of scenario bit-string is \tone, \ttwo, \tonec, \ttwof. For instance, the string 0011 indicates that sequences \tone and \ttwo were synthesized from \tonec and \ttwof sequences.} Patient images shown here are from \emph{Brats18\_CBICA\_AAP\_1} from HGG cohort.}
    \label{fig:bratsqual}
\end{figure*}
% -------------------------------------------------------------------------------
% -------------------------------------------------------------------------------

% \begin{figure*}[h]
%     \centering
%     \includegraphics[height=0.95\textheight]{figs/pdf/exp1qualt1t2pat2.pdf}
%     \caption{Qualitative results for \pptone, \pgtone, and \mitone. Top two and bottom two subfigures visualize two patient's axial slices for sequences \tone and \ttwo, while showing synthesized slices from \pptone, \pgtone and \mitone models. Red arrows show areas where tumor regions were present, and successfully synthesized by \mi in both \tone and \ttwo synthesis.}
%     \label{fig:exp1qualt1}
% \end{figure*}

% Table generated by Excel2LaTeX from sheet 'Sheet1'
\begin{table}[h]
  \centering
  \caption{Comparison between \pp, \pg and \mi. Values in boldface represent best performance values. Reported values are mean$\pm$std.}
    \begin{tabular}{lrrr}
    \toprule
    \multicolumn{1}{c}{\textbf{Model}} & \multicolumn{1}{c}{\textbf{MSE}} & \multicolumn{1}{c}{\textbf{PSNR}} & \multicolumn{1}{c}{\textbf{SSIM}} \\
    \midrule
    \pptone & 0.0135$\pm$0.0044 & 22.1168$\pm$2.1001 & 0.8864$\pm$0.0180 \\
    \pgtone & 0.0107$\pm$0.0048 & 23.8645$\pm$2.8851 & 0.8992$\pm$0.0203 \\
    \mitone & \textbf{0.0052$\pm$0.0026} & \textbf{26.6057$\pm$1.3801} & \textbf{0.9276$\pm$0.0118} \\
    \midrule
    \ppttwo & 0.0050$\pm$0.0019 & 25.0606$\pm$1.2020 & 0.8931$\pm$0.0176 \\
    \pgttwo & 0.0050$\pm$0.0033 & 25.4511$\pm$1.6773 & 0.9008$\pm$0.0250 \\
    \mittwo & \textbf{0.0049$\pm$0.0041} & \textbf{26.1233$\pm$2.6630} & \textbf{0.9078$\pm$0.0324} \\
    \bottomrule
    \end{tabular}%
  \label{tab:comp_pgan}%
\end{table}%

% \begin{figure*}[t]
%     \centering
%     \includegraphics[height=0.45\textheight]{figs/pdf/exp1qual1t2_2.pdf}
%     \caption{\red{Qualitative results for \pp, \pg, and \mi. Left subfigure (columns (a) through  (d)) visualize the performance of three models in synthesis of \tone sequence. The second subfigure (columns (e) through (h)) show synthesized \ttwo sequences for each of the tested models. Red arrows indicate areas where tumor regions were present, and successfully synthesized by \mi in both \tone and \ttwo synthesis.}}
%     \label{fig:exp1qualt1}
% \end{figure*}

\red{Table \ref{tab:comp_pgan} presents MSE, PSNR and SSIM results for all three model architectures and their variants. We observe that both variants of \mi outperform both \pp and \pg models in all metrics.}

\red{Comparing \mitone and the baseline \pptone, \mitone outperformed by 61.48\% in terms of MSE ($p<$ 0.05), 20.29\% in PSNR ($p<$ 0.05) and 4.64\% in SSIM ($p<$ 0.05). \mitone also outperformed the state-of-art single-input single-output model \pgtone, in all metrics, with improvements of 51.40\% in MSE ($p<$ 0.01), 11.48\% in PSNR ($p<$ 0.05) and 3.15\% in SSIM ($p<$ 0.05).}

\red{Similarly for \ttwo synthesis, \mittwo outperforms both \ppttwo, \pgttwo. With respect to \ppttwo, \mittwo performs better by 2\% in MSE, 4.24\% in PSNR and 1.64\% in terms of SSIM. Compared to \pgttwo, \mittwo shows improvement of 2\% in MSE, 2.64\% in PSNR and 0.77\% in SSIM.}

% \red{Similarly for \ttwo synthesis, \mittwo outperforms both \ppttwo, \pgttwo. With respect to \ppttwo, \mittwo performs better by 2\% in MSE (0.0049 $\pm$ 0.0041 vs 0.0050 $\pm$ 0.0019), 4.24\% in PSNR (26.1233 $\pm$ 2.6630 vs 25.0606 $\pm$ 1.2020) and 1.64\% in terms of SSIM (0.9078 $\pm$ 0.0324 vs 0.8931 $\pm$ 0.0176). Compared to \pgttwo, \mittwo shows improvement of 2\% in MSE (0.0049 $\pm$ 0.0041 vs 0.0050 $\pm$ 0.0033), 2.64\% in PSNR (26.1233 $\pm$ 2.6630 vs 25.4511 $\pm$ 1.6773) and 0.77\% in SSIM (0.9078 $\pm$ 0.0324 vs 0.9008 $\pm$ 0.0250).}

\red{These improvements of \mi over \pp and \pg models can be attributed to the availability of multiple sequences as input, which the network utilizes to synthesize missing sequences. The qualitative results showing axial slices from a test patient are provided in Suppl. Mat. Figure S1 in which red arrow points to the successful synthesis of tumor regions in the case of \mi, which was possible due to tumor specific information present in the available three sequences about the various tumor sub-regions (edema, enhancing and necrotic core) in the input sequences, which is not available in its entirety to the single-input single-output methods. We also notice that \mi performs consistently for a single patient, without showing significant deviation from the ground truth intensity distributions in both \tone and \ttwo.}

\red{Superior quantitative and qualitative results showing \mi outperforming \pp and \pg reinforce the hypothesis that using multiple input sequences for the synthesis of a missing sequence is objectively better than using just one input sequence. Moreover, using multi-input methods reduces the number of required models by an order of magnitude, where for a multi-input single-output (many-to-one) only 4 models would be required, compared to 12 for single-input single-output (one-to-one) model ($C(C-1)$ when $C =$ number of sequences $=4$). A multi-input multi-output (many-to-many) model which we explore in this work, improves this further by just requiring a single model to perform all possible synthesis tasks for a given $C$, leading to enormous computational savings during training time.}

\subsection{\ttwof Synthesis (MISO)}
\label{subsec:multimodal}
In this second set of experiments we train our MM-GAN model to synthesize \ttwof sequence in the presence of a varied number of input sequences (one, two or three). Contrasting from the MI-GAN models, this model is trained to generalize on number of different scenarios depending on the available input sequences. In this case, the number of valid scenarios are 7. We perform validation on the ISLES2015 dataset in order to directly compare with  REPLICA \cite{Jog2017_REPLICA} and MM-Synthesis \cite{Chartsias2018}. The quantitative results are given in Table \ref{tab:comparison}. We note that the proposed MM-GAN (0.226$\pm$0.046) clearly outperforms REPLICA's unimodal synthesis models (0.271$\pm$0.10) in all scenarios, as well as MM-Synthesis (0.236$\pm$0.08) in majority (4/7) scenarios. Our method also demonstrates an overall lower MSE standard deviation throughout testing (ranging between [0.03, 0.07], compared to REPLICA [0.08, 0.16] and MM-Synthesis [0.02, 0.13]) in all scenarios but one (\ttwo missing). The qualitative results for ISLES2015 are shown in Figure \ref{fig:islesqual}. Compared to MM-Synthesis (from qualitative results shown in their original paper \cite{Chartsias2018}), our results are objectively sharper, with lower blurring artifacts. MM-GAN also preserves high frequency details of the synthesized sequence, while MM-Synthesis and REPLICA seem to miss most of these details. \red{We request the readers to refer to the original MM-Synthesis~\cite{Salman2019_TMI} manuscript’s Figures 5 and 6 for comparison with our proposed MM-GAN's qualitative results given in Figure \ref{fig:bratsqual} of the current manuscript}. Qualitatively from Figure~\ref{fig:islesqual}, MM-GAN follows the intensity distribution of the real \ttwof sequence in its synthesized version of \ttwof. 

We found that using CL based learning did not help in MISO experiments, as the presence of more sequences does not necessarily increase the amount of information available. 
For example, the presence of both \tone and \ttwo does not result in better \ttwof synthesis (MSE 0.2541) compared to  the presence of DW alone (MSE 0.2109).
%in the case of \ttwof synthesis, as was evident from case where the method did worse when both \tone and \ttwo sequence were present (0.2541), than when only DWI was present (0.2109).
This is because, for every missing sequence, there tends to be some ``highly informative" sequences that, if absent, reduces the synthesis performance by a larger margin. On the other hand, the presence of these highly informative sequences can dramatically boost performance, even in cases where no other sequence is present. 
Due to this, the assumption that leveraging a higher number of present sequences implies an easier case (i.e. more accurate synthesis) does not hold, and thus it becomes problematic to rank scenarios based on how easy they are, which renders CL useless in this case. 
% does not hold too well with a very particular scenario of synthesizing \ttwof sequence. 
Globally (for all valid scenarios, presented in next subsection), however, this assumption tends to hold due to the complex nature of interactions between sequences. 
% Ordering scenarios based on ease of synthesis purley  based on the number of sequences present 
CL helps tremendously in achieving a stable training of the network in the subsequent experiments (MIMO). For MISO, every scenario was shown to the network with uniform probability, throughout training. 
% Moreover, training to synthesize just one sequence \ttwof was an easier task than training to synthesize multiple. Since the assumption of more sequences providing more information didn't hold in this case, we resorted to randomly sampling the scenarios with a uniform probability during training.  

% -------------------------------------------------------------------------------
% -------------------------------------------------------------------------------
% Table generated by Excel2LaTeX from sheet 'Sheet1'
\begin{table}[h]
  \centering
  \caption{Comparison with unimodal method REPLICA and multimodal method MM-Synthesis. The reported values are mean squared error (MSE). Boldface values represent lowest values of the three methods for a particular scenario. }
    \begin{tabular}{lccc}
    \toprule
    \textbf{Scenarios} & \textbf{REPLICA} & \textbf{MM-Synthesis} & \textbf{MM-GAN} \\
    \tiny \tone \ttwo DW& & & (Proposed)\\
    \midrule
    -\hspace{3.0mm}-\hspace{3.0mm}\checkmark  & 0.278$\pm$0.09 & 0.285$\pm$0.13 & \textbf{0.210$\pm$0.057} \\
    -\hspace{3.0mm}\checkmark\hspace{2.0mm}-  & 0.374$\pm$0.16 & 0.321$\pm$0.12 & \textbf{0.279$\pm$0.055} \\
    -\hspace{3.0mm}\checkmark\hspace{1.5mm}\checkmark  & 0.235$\pm$0.08 & 0.214$\pm$0.09 & \textbf{0.182$\pm$0.033} \\
    \checkmark\hspace{2.0mm}-\hspace{3.0mm}-  & 0.301$\pm$0.11 & \textbf{0.249$\pm$0.09} & 0.281$\pm$0.071 \\
    \checkmark\hspace{2.0mm}-\hspace{2.5mm}\checkmark  & 0.225$\pm$0.08 & 0.198$\pm$0.02 & \textbf{0.191$\pm$0.039} \\
    \checkmark\hspace{1.5mm}\checkmark\hspace{2.0mm}-  & 0.271$\pm$0.12 & \textbf{0.214$\pm$0.08} & 0.254$\pm$0.066 \\
    \checkmark\hspace{1.5mm}\checkmark\hspace{1.5mm}\checkmark  & 0.210$\pm$0.08 & \textbf{0.171$\pm$0.06} & 0.182$\pm$0.041\\
    \midrule
    \textbf{Mean} & 0.271$\pm$0.10 & 0.236$\pm$0.08 & \textbf{0.226$\pm$0.046} \\
    \bottomrule
    \end{tabular}%
  \label{tab:comparison}%
\end{table}%

\subsection{Multimodal Synthesis (MIMO)}
\label{subsec:mimo}
We present results for our experiments on BRaTS2018's HGG and LGG cohorts in Table \ref{tab:brats2018hgg} \red{and \ref{tab:brats2018lgg}}. \red{We set $z=$ zeros for imputation, and train the networks with implicit conditioning (IC) and curriculum learning (CL)}. In this experiment we train our proposed MM-GAN model on all 14 valid scenarios, in order to synthesize any missing sequence from any number or combination of available sequences. We observe that the proposed MM-GAN model performs consistently well when synthesizing just one sequence, with high overall SSIM ($>$0.90 in most cases except one in LGG), \red{PSNR ($>$22.0)} values, and low \red{MSE ($<$0.015)}. As more sequences start missing, the task of synthesizing missing sequences gets harder. During the initial epochs of training, MM-GAN tends to learn the general global structure of the brain, without considering the local level details. This seems to be enough for the generator to fool the discriminator initially. However, as the training progresses and the discriminator becomes stronger, the generator is forced to learn the local features of the slice, which includes small details, especially the boundaries between the grey and white matter visible in the sequence. The qualitative results shown in Figure \ref{fig:bratsqual} show how MM-GAN effectively synthesizes the missing sequence in various scenarios, while preserving high frequency details that delineate between grey and white matter of the brain, as well as recreating the tumor region in the frontal lobe by combining information from available sequences. The synthesis of the tumor in the final images depend heavily on the available sequences. For example, the contrast sequence \tonec provides clear delineation of enhancing ring-like region around the necrotic mass, which is an important indicator of the size of the tumor.
Presence of \tone and/or \ttwof sequence leads to improved synthesis of edema features. The contrast sequence \tonec provides unique information about the enhancing region around the tumor, which is usually not visible in any other sequence. Qualitatively, the \ttwo sequence does not seem to directly aid in synthesizing a particular region of tumor well, but coupled with other available sequences, it helps in better synthesis of tumor mass in the final synthesized slice (Figure \ref{fig:bratsqual}). 

% Table generated by Excel2LaTeX from sheet 'Sheet1'
\begin{table}[h]
  \centering
  \caption{Performance on BraTS2018 High Grade Glioma (HGG) Cohort}
    \begin{tabular}{llll}
    \toprule
    \textbf{Scenarios} &       &       &  \\
          \tiny \tone \ttwo \tonec $T_{2f}$& \multicolumn{1}{c}{\textbf{MSE}} & \multicolumn{1}{c}{\textbf{PSNR}} & \multicolumn{1}{c}{\textbf{SSIM}} \\
    \midrule
    -\hspace{3mm}-\hspace{3mm}-\hspace{3mm}\checkmark  & 0.0143$\pm$0.0086 & 23.196$\pm$4.2908 & 0.8973$\pm$0.0668 \\
     -\hspace{3.0mm}-\hspace{3.0mm}\checkmark\hspace{2mm}- & 0.0072$\pm$0.0065 & 24.524$\pm$4.0671 & 0.8984$\pm$0.0726 \\
    -\hspace{3.0mm}-\hspace{3.0mm}\checkmark\hspace{1.5mm}\checkmark & 0.0060$\pm$0.0061 & 25.863$\pm$3.2218 & 0.9166$\pm$0.0339 \\
    -\hspace{3.0mm}\checkmark\hspace{2.0mm}-\hspace{3.0mm}- & 0.0102$\pm$0.0065 & 23.469$\pm$4.1744 & 0.9074$\pm$0.0680 \\
    -\hspace{3.0mm}\checkmark\hspace{2.0mm}-\hspace{2.5mm}\checkmark  & 0.0136$\pm$0.0048 & 22.900$\pm$2.1989 & 0.9156$\pm$0.0260 \\
    -\hspace{3.0mm}\checkmark\hspace{1.5mm}\checkmark\hspace{2.0mm}- & 0.0073$\pm$0.0070 & 24.792$\pm$2.9524 & 0.9140$\pm$0.0311 \\
    -\hspace{3.0mm}\checkmark\hspace{1.5mm}\checkmark\hspace{1.5mm}\checkmark & 0.0091$\pm$0.0053 & 24.173$\pm$3.2754 & 0.9228$\pm$0.0190 \\
    \checkmark\hspace{2.0mm}-\hspace{3.0mm}-\hspace{3.0mm}- & 0.0072$\pm$0.0056 & 24.879$\pm$3.8216 & 0.9091$\pm$0.0651 \\
    \checkmark\hspace{2.0mm}-\hspace{3.0mm}-\hspace{2.5mm}\checkmark & 0.0073$\pm$0.0041 & 26.189$\pm$2.1337 & 0.9264$\pm$0.0328 \\
    \checkmark\hspace{2.0mm}-\hspace{2.5mm}\checkmark\hspace{2.0mm}- & 0.0040$\pm$0.0032 & 26.150$\pm$1.8470 & 0.9107$\pm$0.0275 \\
    \checkmark\hspace{2.0mm}-\hspace{2.5mm}\checkmark\hspace{1.5mm}\checkmark & 0.0017$\pm$0.0026 & 28.678$\pm$2.3290 & 0.9349$\pm$0.0262 \\
    \checkmark\hspace{1.5mm}\checkmark\hspace{2.0mm}-\hspace{3.0mm}- & 0.0068$\pm$0.0041 & 25.242$\pm$2.0339 & 0.9175$\pm$0.0275 \\
    \checkmark\hspace{1.5mm}\checkmark\hspace{2.0mm}-\hspace{2.5mm}\checkmark & 0.0098$\pm$0.0066 & 24.372$\pm$2.2792 & 0.9239$\pm$0.0375 \\
    \checkmark\hspace{1.5mm}\checkmark\hspace{1.5mm}\checkmark\hspace{2mm}- & 0.0033$\pm$0.0040 & 26.397$\pm$1.9733 & 0.9150$\pm$0.0275 \\
    \midrule
    mean$\pm$std & 0.0082$\pm$0.0054 & 24.789$\pm$2.8999 & 0.9120$\pm$0.0401 \\
    \bottomrule
    \end{tabular}%
  \label{tab:brats2018hgg}%
\end{table}%

% Table generated by Excel2LaTeX from sheet 'Sheet1'
\begin{table}[h]
  \centering
  \caption{Performance on BraTS2018 Low Grade Glioma (LGG) Cohort}
    \begin{tabular}{llll}
    \toprule
    \textbf{Scenarios} &       &       &  \\
    \tiny \tone \ttwo \tonec $T_{2f}$ & \multicolumn{1}{c}{\textbf{MSE}} & \multicolumn{1}{c}{\textbf{PSNR}} & \multicolumn{1}{c}{\textbf{SSIM}} \\
    \midrule
    -\hspace{3mm}-\hspace{3mm}-\hspace{3mm}\checkmark  & 0.0092$\pm$0.0037 & 24.7832$\pm$3.3197 & 0.8758$\pm$0.0651 \\
     -\hspace{3.0mm}-\hspace{3.0mm}\checkmark\hspace{2mm}- & 0.0072$\pm$0.0032 & 25.7012$\pm$2.9797 & 0.8925$\pm$0.0486 \\
    -\hspace{3.0mm}-\hspace{3.0mm}\checkmark\hspace{1.5mm}\checkmark & 0.0040$\pm$0.0022 & 26.9985$\pm$2.4075 & 0.9179$\pm$0.0178 \\
    -\hspace{3.0mm}\checkmark\hspace{2.0mm}-\hspace{3.0mm}- & 0.0108$\pm$0.0044 & 23.6527$\pm$3.4296 & 0.8811$\pm$0.0560 \\
    -\hspace{3.0mm}\checkmark\hspace{2.0mm}-\hspace{2.5mm}\checkmark  & 0.0129$\pm$0.0054 & 22.9238$\pm$2.6554 & 0.8873$\pm$0.0337 \\
    -\hspace{3.0mm}\checkmark\hspace{1.5mm}\checkmark\hspace{2.0mm}- & 0.0084$\pm$0.0036 & 24.7581$\pm$2.2131 & 0.8984$\pm$0.0144 \\
    -\hspace{3.0mm}\checkmark\hspace{1.5mm}\checkmark\hspace{1.5mm}\checkmark & 0.0061$\pm$0.0035 & 25.9841$\pm$2.1926 & 0.9288$\pm$0.0137 \\
    \checkmark\hspace{2.0mm}-\hspace{3.0mm}-\hspace{3.0mm}- & 0.0120$\pm$0.0063 & 23.6018$\pm$3.8153 & 0.8908$\pm$0.0509 \\
    \checkmark\hspace{2.0mm}-\hspace{3.0mm}-\hspace{2.5mm}\checkmark & 0.0109$\pm$0.0040 & 23.8408$\pm$2.1715 & 0.9028$\pm$0.0244 \\
    \checkmark\hspace{2.0mm}-\hspace{2.5mm}\checkmark\hspace{2.0mm}- & 0.0102$\pm$0.0030 & 23.9202$\pm$2.0746 & 0.8792$\pm$0.0178 \\
    \checkmark\hspace{2.0mm}-\hspace{2.5mm}\checkmark\hspace{1.5mm}\checkmark & 0.0057$\pm$0.0046 & 25.6005$\pm$2.7909 & 0.9030$\pm$0.0303 \\
    \checkmark\hspace{1.5mm}\checkmark\hspace{2.0mm}-\hspace{3.0mm}- & 0.0128$\pm$0.0048 & 22.1330$\pm$1.8389 & 0.8885$\pm$0.0164 \\
    \checkmark\hspace{1.5mm}\checkmark\hspace{2.0mm}-\hspace{2.5mm}\checkmark & 0.0120$\pm$0.0025 & 22.4980$\pm$1.1684 & 0.9086$\pm$0.0253 \\
    \checkmark\hspace{1.5mm}\checkmark\hspace{1.5mm}\checkmark\hspace{2mm}- & 0.0113$\pm$0.0040 & 23.0852$\pm$1.5142 & 0.8692$\pm$0.0224 \\
    \midrule
    mean$\pm$std  & 0.0095$\pm$0.0039 & 24.2487$\pm$2.4694 & 0.8946$\pm$0.0312 \\
    \bottomrule
    \end{tabular}%
  \label{tab:brats2018lgg}%
\end{table}%

% \begin{figure}[h]
%     \centering
%     \includegraphics[width=0.45\textwidth]{figs/pdf/performance_cropped.pdf}
%     \caption{Performance degradation as number of sequences missing increase. The figure shows that the method degrades elegantly, with no sharp drops.}
%     \label{fig:performanceDrop}
% \end{figure}
\null\vspace{-0.5cm}
\begin{figure}[h]
    \centering
    \includegraphics[width=0.34\textwidth]{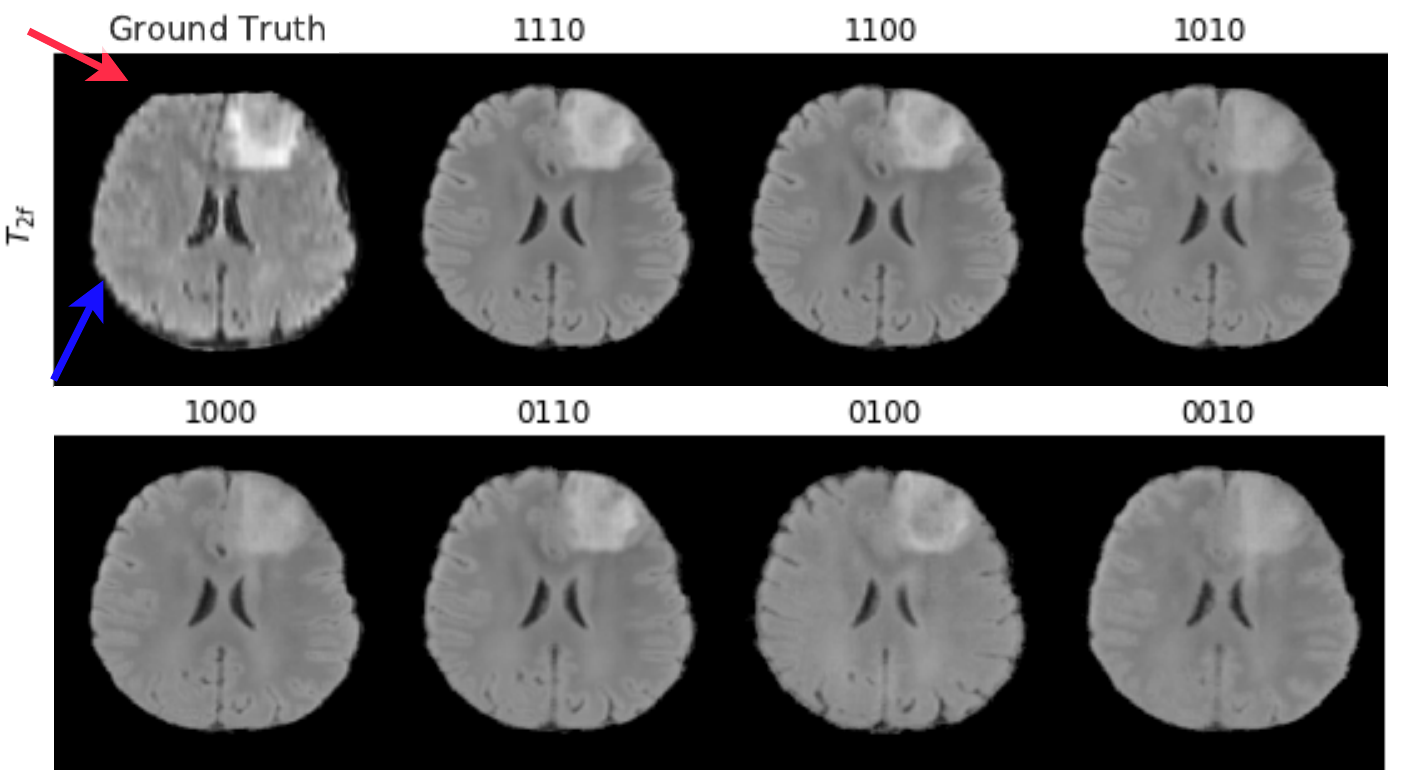}
    \caption{Illustration of MM-GAN filling up parts in the scan that are originally missing in the ground truth. Red arrow shows the top part of brain that is missing, and is synthesized in all scenarios. Blue arrow highlights the high-frequency details that are missing in ground truth, but are synthesized in most images. \red{We also notice that when all sequences are present, the \ttwof synthesis is the most accurate with respect to the pathology visible in the frontal cortex.} The order of scenario bit-string is \tone, \ttwo, \tonec, \ttwof. Patient images shown here are from \emph{Brats18\_2013\_9\_1} from LGG cohort.}
    \label{fig:show_viz}
\end{figure}

As shown in Figure \ref{fig:show_viz}, we also observe that the method fills up lost details as can be seen in \ttwof sequence. The original ground truth sequence has the frontal lobe part cut off, probably due to patient movement or miss-registration. However MM-GAN recreates that part by using information from the available sequences. Another interesting side-effect of our approach is visible in \ttwof synthesis, where the synthesized versions of \ttwof exhibit higher quality details (Figure \ref{fig:show_viz}) than the original sequence, which was acquired in a very low resolution. This effect is the consequence of the method using high-resolution input sequences (all sequences except \ttwof are acquired at higher resolution) to synthesize the missing \ttwof sequence. This also suggests that our method may be used for improving or upscaling resolution of available sequences. However we do not investigate this further here, and leave it as future work. We found that normalizing sequences with mean value is easier to train with, and naturally supports final layer activations like ReLU.
% We also investigated zero mean and unit variance normalization and found that it under-performs relative to the mean normalization. 

% Table generated by Excel2LaTeX from sheet 'Sheet3'
\begin{table}[htbp]
  \centering
    \caption{MM-GAN Performance variation with respect to number of sequences missing for HGG and LGG cohort.}
    \begin{tabular}{@{\hskip2pt}c@{\hskip3.1pt}c@{\hskip3.1pt}c@{\hskip3.1pt}c@{\hskip3.1pt}c}
    \toprule
    \textbf{Dataset} & \textbf{Missing} & \textbf{MSE} & \textbf{PSNR} & \textbf{SSIM} \\
    \midrule
    \multirow{3}[2]{*}{HGG} & 1     & 0.0539 $\pm$ 0.0215 & 29.1162 $\pm$ 0.9716 & 0.9268 $\pm$ 0.0115 \\
          & 2     & 0.0602 $\pm$ 0.0170 & 28.9064 $\pm$ 0.8095 & 0.9223 $\pm$ 0.0090 \\
          & 3     & 0.0752 $\pm$ 0.0125 & 28.0801 $\pm$ 0.4110 & 0.9087 $\pm$ 0.0071 \\
\cmidrule{2-5}    \multirow{3}[2]{*}{LGG} & 1     & 0.1296 $\pm$ 0.0485 & 26.1362 $\pm$ 2.8872 & 0.9080 $\pm$ 0.0383 \\
          & 2     & 0.1499  $\pm$ 0.0208 & 25.5641 $\pm$ 1.3161 & 0.8976 $\pm$ 0.0157 \\
          & 3     & 0.1914 $\pm$ 0.0358 & 24.8987 $\pm$ 0.8363 & 0.8732 $\pm$ 0.0205 \\
    \bottomrule
    \end{tabular}%
  \label{tab:performanceDrop}%
\end{table}%

% \begin{table}[h]
%   \centering
%   \caption{MM-GAN Performance variation with respect to number of sequences missing for HGG (top) and LGG (bottom) cohort.}
%     \begin{tabular}{cccc}
%     \toprule
%     \multicolumn{1}{c}{\textbf{Missing}} & \textbf{MSE} & \textbf{PSNR} & \textbf{SSIM} \\
%     \midrule
%     1     & 0.0539 $\pm$ 0.0215 & 29.1162 $\pm$ 0.9716 & 0.9268 $\pm$ 0.0115 \\
%     2     & 0.0602 $\pm$ 0.0170 & 28.9064 $\pm$ 0.8095 & 0.9223 $\pm$ 0.0090 \\
%     3     & 0.0752 $\pm$ 0.0125 & 28.0801 $\pm$ 0.4110 & 0.9087 $\pm$ 0.0071 \\
%     \bottomrule
%     \end{tabular}%
%   \label{tab:perfhgg}%
% \end{table}%

% \null\vspace{-0.95cm}
% % LGG
% % Table generated by Excel2LaTeX from sheet 'Sheet3'
% \begin{table}[h]
%   \centering
%     \begin{tabular}{cccc}
%     \toprule
%     \multicolumn{1}{c}{\textbf{Missing}} & \textbf{MSE} & \textbf{PSNR} & \textbf{SSIM} \\
%     \midrule
%     1     & 0.1296 $\pm$ 0.0485 & 26.1362 $\pm$ 2.8872 & 0.9080 $\pm$ 0.0383 \\
%     2     & 0.1499 $\pm$ 0.0208 & 25.5641 $\pm$ 1.3161 & 0.8976 $\pm$ 0.0157 \\
%     3     & 0.1914 $\pm$ 0.0358 & 24.8987 $\pm$ 0.8363 & 0.8732 $\pm$ 0.0205 \\
%     \bottomrule
%     \end{tabular}%
%   \label{tab:perflgg}%
% \end{table}%

We observe that the generators perform really well when they are constrained using a non-linear activation function at the final layer. However in the case of MISO, the limitation of the normalization type (dividing by mean value of sequence) used in MM-Synthesis prevents us from using a non-linear activation at the end of generator. This is due to the fact that some patients' data in SISS cohort from ISLES2015 contain negative intensity values, which after normalization stay negative. It can be seen that the MSE values reported in Table \ref{tab:comparison} tend to be higher than the ones reported in \red{Tables \ref{tab:brats2018hgg} and \ref{tab:brats2018lgg}}, due to the latter set of experiments using ReLU activation at the end of generator. 

% R. 1.1 comment
\red{Although MM-GAN observes different scenarios and hence different fake sequences in each iteration, which may affect stability during training, we did not observe any unstable behaviour during the training process. The use of implicit conditioning (IC) assisted in ensuring stable training of networks by making the task challenging for the discriminator, preventing it from overpowering the generator, which in turn lead to the generator converging to a good local minima. }

We also observe that the proposed method shows graceful degradation as the number of sequences missing start increasing, which is apparent both qualitative and quantitatively in Figure \ref{fig:bratsqual} and Table \ref{tab:performanceDrop}. \red{For instance, in HGG experiments, compared to having one sequence missing, the performance of MM-GAN drops on average by 27.1\%, 2.7\% and 0.7\% in MSE, PSNR and SSIM respectively for scenarios where two sequences are missing. For scenarios where three sequences are missing, the performance drops on average by 39.1\%, 7.2\% and 2.2\% in terms of MSE, PSNR and SSIM respectively compared to one sequence missing, and 29.3\%, 4.6\% and 1.5\% when compared to scenarios where two sequences are missing. We observe that the method holds up well in generating sequences with high fidelity in terms of PSNR and SSIM even in harder scenarios where multiple sequences may be missing.}

\red{Qualitatively, we also investigated the question as to which sequences are the most valuable for the synthesis for each of the four sequences in BraTS2018 HGG cohort. For every sequence that is synthesized, we list a ranking based upon our investigation of the results for each of the remaining sequences. For synthesizing \ttwof, we found that \tonec sequence, followed by \ttwo and \tone sequences were important. This is also apparent in Figure~\ref{fig:bratsqual}, where the removal of \tonec in column (c) lead to the synthesized  sequence missing necrotic part of tumor completely, while the removal of \tone (columns (b) and (f)) and \ttwo (columns (b) and (d)) did not affect the performance dramatically. For the synthesis of \tonec, we found that \ttwo sequence held the highest significance, followed by \tone and \ttwof (comparing columns (b) with (d), (f), (c)). This is also evident from row 3 column (d) in Figure 3, when removal of \ttwo sequence lead to increased blurring artifacts in the synthesized version of \tonec, which were not as pronounced when \tone or \ttwof were removed. For \ttwo synthesis, we found that \tonec sequence contributed the most towards accurate synthesis (comparing columns (b) and (d)), with \tone sequence also playing an important role (columns (b) and (f)), lastly followed by \ttwof. Finally for \tone synthesis, we found that \tonec was the most important sequence (columns (b) and (d)) enabling accurate synthesis, followed closely by \ttwo (columns (b) and (f)) and \ttwof (columns (b) and (c)).}

\red{Due to MM-GAN being a single unified model (MM-GAN), it relieves the end-user from the difficult task of choosing the right model for synthesis during inference time. For instance, in the case where sequences (\tone,\ttwo,\tonec) are present, and \ttwof to be synthesized, a single-input single-output method would have three networks capable of synthesizing \ttwof from \tone, \ttwo and \tonec respectively. The decision as to which network should be chosen for this problem is hard, since each unimodal network would provide trade-offs in terms synthesis quality, especially in tumorous areas where individual sequences do not provide full information. This decision problem is mitigated in multi-input models (MM-Synthesis and MI-GAN), but there still exists the computational overhead during training time in order to train multiple models for each output sequence (total 4 for both MM-Synthesis and MI-GAN). MM-GAN on the other hand, is completely multimodal and only requires training for just one model, which provide computational savings during training time by eliminating the need for training multiple models (if number of sequences $C$=4, then 12 models in case of unimodal, 4 models in case of multi-input multi-output architectures).}

\section{Conclusion}
\label{sec:conclusion}
We propose a multi-modal generative adversarial network (MM-GAN) capable of synthesizing missing MR pulse sequences using combined information from the available sequences. Most approaches so far in this domain had been either unimodal, or partially multi-modal (multi-input, single-output). We present a truly multi-modal method that is multi-input and multi-output, generalizing to any combination of available and missing sequences. The synthesis process runs in a single forward pass of the network regardless of the number of sequences missing, and the run time is constant w.r.t number of missing sequences. 

The first variant of our proposed MM-GAN, called MI-GAN outperformed the unimodal version pGAN in all three metrics (Table \ref{tab:comp_pgan}). We also show that MM-GAN outperforms the best multimodal synthesis method REPLICA~\cite{Jog2017_REPLICA}, as well as MM-Synthesis~\cite{Chartsias2018} in multi-input single-input synthesis of \ttwof sequence (Table \ref{tab:comparison}), and produces objectively sharper and more accurate results. In another set of experiments, we train our method on BraTS2018 dataset to set up a new benchmark in terms of MSE, PSNR and SSIM (\red{Tables \ref{tab:brats2018hgg} and \ref{tab:brats2018lgg}}), and show qualitative results for the same (Figure \ref{fig:bratsqual}). MM-GAN performance degrades as a function of number of missing sequences in Table \ref{tab:performanceDrop} but exhibits robustness in maintaining high PSNR and SSIM values even in harder scenarios. Finally, we show that our method is capable of filling in details missing from the original ground truth sequences, and also capable of improving quality of the synthesized sequences (Figure \ref{fig:show_viz}). 

Although our approach qualitatively and quantitatively performs better than all other competing methods, we note that it has problems in synthesizing the enhancing subregion in \tonec sequence properly. This, however, is expected since \tonec sequence contains highly specific information about the enhancing region of the tumor that is not present in any other sequences. \red{An inherent limitation of all synthesis methods stems from the fact that MR sequences provide both redundant and unique information. This creates challenges for all synthesis methods, unimodal and multimodal alike. Unimodal methods provide a one-to-one mapping between sequences, but each such model (12 total for 4 sequences) would raise tradeoffs between the synthesis accuracy. For instance, in the experiments given in Section VI.B, we found that, \redtwo{in terms of overall sequence synthesis performance}, synthesizing \ttwof from DWI tends to be more accurate (MSE 0.2109) than synthesizing from \tone (MSE 0.2813)  or \ttwo (MSE 0.2799). This reinforces the fact that there are some inherent characteristics to each sequence, which can only be faithfully synthesized if another sequence that more or less captures similar characteristics is present. The sequences provide complementary visual information for a human reader, though there are underlying correlations imperceptible to the naked eye, since they all originate due to common underlying physics and from the same subject. Multi-input methods like ours can exploit the correlations between available sequences, and synthesize the missing sequence by leveraging information from all input sequences. This is evident from the quantitative results in \red{Tables \ref{tab:comparison}, \ref{tab:brats2018hgg}, \ref{tab:brats2018lgg}, \ref{tab:performanceDrop}}, and summarized in Table \ref{tab:performanceDrop} where more available sequences allow better synthesis of missing ones}. For future work, we note that the inherent design of our method is 2D, and an extension of the work which can take either 2.5D or 3D images into account may perform better both quantitatively and qualitatively. 
\red{Another area of investigation would be to explore the up-scaling capabilities of the MM-GAN, where given a low-quality ground truth scan with missing scan areas, the method can generate a higher quality version with filled in missing details. It would also be interesting to test MM-GAN by deploying it as part of the pipeline for downstream analysis, for example segmentation. This natural placement in the pipeline would allow the downstream methods to become robust to missing pulse sequences. Compared to HeMIS~\cite{Havaei2016_HeMIS} and PIMMS~\cite{Varsavsky2018_PIMSS}, this can be another approach to make segmentation algorithms robust to missing pulse sequences.  }

\section{Acknowledgements}
\label{sec:ack}
 The authors would like to thank NVIDIA Corporation for donating a Titan X GPU. This research was enabled in part by support provided by WestGrid (www.westgrid.ca) and Compute Canada (www.computecanada.ca). \red{We  thank the anonymous reviewers for their insightful feedback that resulted in a much improved paper. }

\bibliographystyle{IEEEtran}  
\bibliography{library}  

% that's all folks
\end{document}

\captionsetup[figure]{format=cancaption,labelformat=cancaptionlabel}
\captionsetup[table]{format=cancaption,labelformat=cancaptionlabel}

\bstctlcite{IEEEexample:BSTcontrol}
% @IEEEtranBSTCTL{IEEEexample:BSTcontrol,
% CTLdash_repeated_names = "no",
% CTLuse_forced_etal       = "yes",
% CTLmax_names_forced_etal = "4",
% CTLnames_show_etal       = "1" }

%
% paper title
% Titles are generally capitalized except for words such as a, an, and, as,
% at, but, by, for, in, nor, of, on, or, the, to and up, which are usually
% not capitalized unless they are the first or last word of the title.
% Linebreaks \\ can be used within to get better formatting as desired.
% Do not put math or special symbols in the title.
\title{Missing MRI Pulse Sequence Synthesis using Multi-Modal Generative Adversarial Network}
\author{Anmol~Sharma,~\IEEEmembership{Student~Member,~IEEE,}
        Ghassan~Hamarneh,~\IEEEmembership{Senior~Member,~IEEE}% <-this % stops a space
\thanks{This work was partially supported by the NSERC-CREATE Bioinformatics 2018-2019 Scholarship.}
\thanks{Anmol Sharma and Ghassan Hamarneh are with the Medical Image Analysis Laboratory, School of Computing Science, Simon Fraser University, Canada. 
e-mail: \{asa224, hamarneh\}@sfu.ca}% <-this % stops a space
}

% The paper headers
\ifCLASSOPTIONpeerreview
    \markboth{Transaction on Medical Imaging}%
    {IEEE TMI}
\else
    % \markboth{IEEE Journal}%
    % {Second Header Target Journal}
    \markboth{Supplementary Material}%
    {}
\fi

\newcommand{\tone}{$T_1$\xspace} 
\newcommand{\tonec}{$T_{1c}$\xspace} 
\newcommand{\ttwo}{$T_2$\xspace} 
\newcommand{\ttwof}{$T_{2flair}$\xspace}
\newcommand{\mmgan}{MM-GAN\xspace}
\newcommand{\allseq}{\tone, \tonec, \ttwo, \ttwof}
\newcommand{\G}{$\mathcal{G}$\xspace}
\newcommand{\D}{$\mathcal{D}$\xspace}

\newcommand{\pptone}{$\text{P2P}_{T_1}$\xspace}
\newcommand{\ppttwo}{$\text{P2P}_{T_2}$\xspace}
\newcommand{\pp}{$\text{P2P}$\xspace}

\newcommand{\pgtone}{$\text{pGAN}_{T_1}$\xspace}
\newcommand{\pgttwo}{$\text{pGAN}_{T_2}$\xspace}
\newcommand{\pg}{$\text{pGAN}$\xspace}

\newcommand{\mitone}{$\text{MI-GAN}_{T_1}$\xspace}
\newcommand{\mittwo}{$\text{MI-GAN}_{T_2}$\xspace}
\newcommand{\mi}{$\text{MI-GAN}$\xspace}
\newcommand{\red}[1]{\textcolor{black}{#1}}
\newcommand{\redtwo}[1]{\textcolor{black}{#1}}
\maketitle
\begin{figure*}[h]
    \centering
    \includegraphics[height=0.45\textheight]{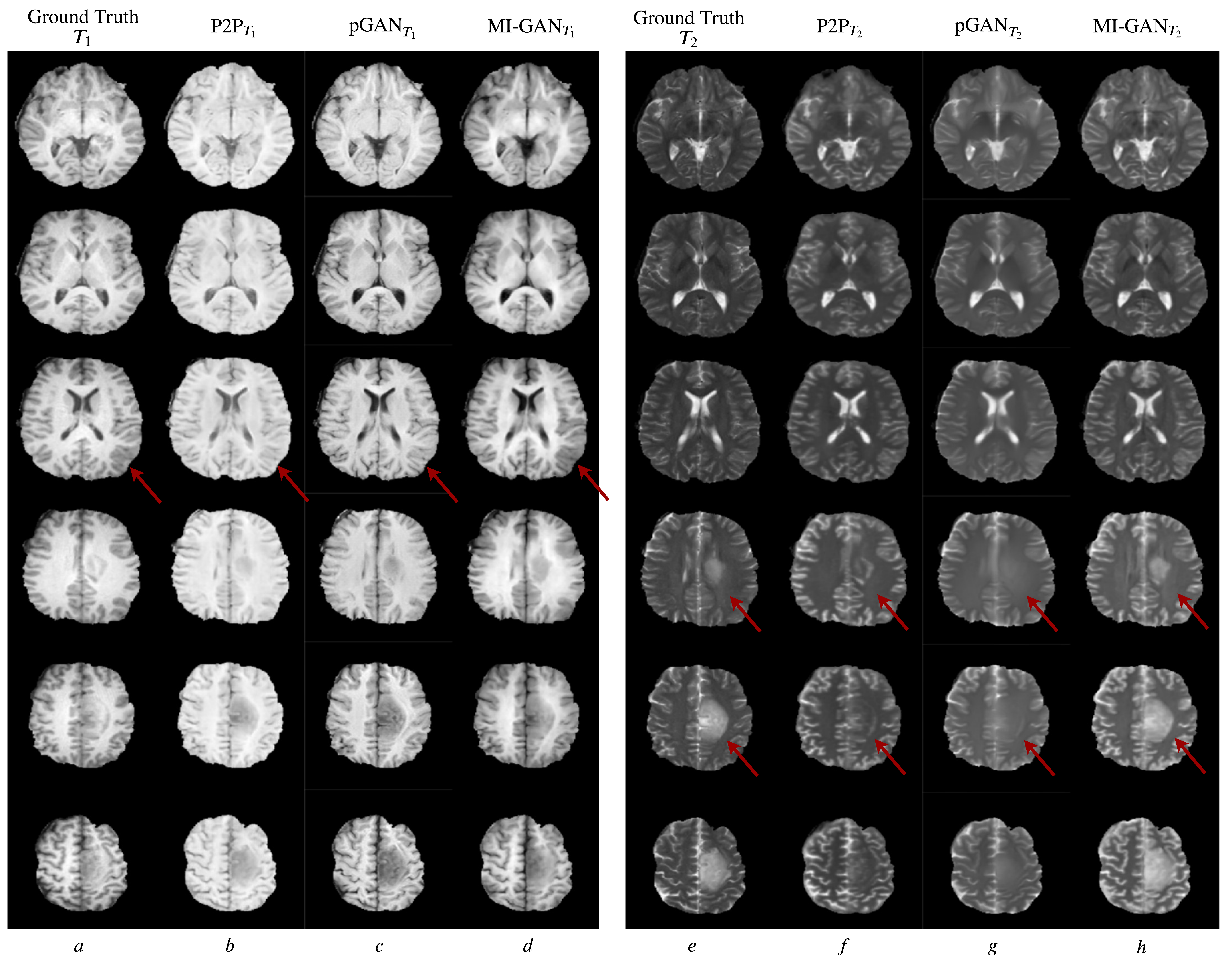}
    \caption{\red{Qualitative results for \pp, \pg, and \mi. Left subfigure (columns (a) through  (d)) visualize the performance of three models in synthesis of \tone sequence. The second subfigure (columns (e) through (h)) show synthesized \ttwo sequences for each of the tested models. Red arrows indicate areas where tumor regions were present, and successfully synthesized by \mi in both \tone and \ttwo synthesis.}}
    \label{fig:exp1qualt1}
\end{figure*}

\section{Validation of Key Components}
\subsection{Imputation Value and Curriculum Learning}
\label{subsubsec:imputation}
\red{In our proposed multi-input multi-output MM-GAN model, the missing sequences during both training and test time has to be imputed with some value. In order to identify the best approach for imputing missing slices, we test \mmgan with three imputation strategies, where  $z=$ average of available sequences, $z=$ noise, and $z=$ zeros. }

\red{We also test curriculum learning (CL) based training strategy where the network is initially shown easier examples followed by increasingly difficult examples as the training progresses. We compare this approach with random sampling (RS) method, where scenarios are uniformly sampled during training regardless of the number of missing sequences. In order to test the effectiveness of CL based learning, we report results with CL-based learning as well as random sampling (RS) based learning for each imputation strategy.}

\red{The results for our experiments are given in Suppl. Mat. Table S\ref{tab:clnoclz}. 
The imputation value $z=$ zeros performed the best when CL based training strategy is adopted (MSE 0.0095 and PSNR 24.2487), and was slightly outperformed in SSIM by $z=$ zeros with RS sampling (SSIM 0.8955). $z=$ noise performs slightly worse than zero imputation, but better than the $z=$ average imputation for both CL and RS strategies. 
This may be due to the fact that imputing with a fixed value (zero vs noise/average) provides network with a clear indication about a missing sequence (implicit conditioning), as well as relieves it from learning weights that have to generalize for variable values imputed at test time.
Imputing with zeros also removes the non-deterministic behaviour that was prevalent in $z=$ noise, where each test run would yield slightly different quantitative (though mostly similar qualitative) results. 
For our final benchmark results on LGG and HGG cohorts in BraTS2018, we set $z=$ zeros and use CL based learning strategy to train the networks. }

\subsection{Effect of Implicit Conditioning (IC)}
\label{subsubsec:icnoic}
\red{We proposed to use a combination of zero imputation, selective loss computation in \G, and selective discrimination in \D, collectively referred to as implicit conditioning (IC) in order to better train the network to learn to synthesize only the missing sequences, without penalizing it for inaccurate synthesis of other available sequences. In this set of experiments we test the effectiveness of IC based training by training an \mmgan model on the LGG cohort, with and without IC enabled. In the model with IC disabled, we penalize the generator for inaccurately synthesizing all sequences (C = 4), as well as let the discriminator take the whole \G output ($G(X_z)$) as input, instead of imputing real sequences in them ($X_i$). We keep the zero imputation part intact due to the network architectural constraints. }

\red{The results are shown in Suppl. Mat. Table S\ref{tab:icnoic}. We observe that training a model with implicit conditioning turned on outperforms the one which is trained without it in all three metrics. IC-based trained model achieves a MSE of 0.0095, compared to 0.0100 for non-IC model. Similarly, IC-based trained model outperforms in PSNR (24.2487 vs 23.6626), and in SSIM (0.8946 vs 0.8796) ($p<$ 0.01). This performance improvement can be attributed to the fact that the generator has to work harder in case where IC is not used, since it has to synthesize all the sequence with high accuracy in order to achieve a lower loss value. This leads to a trade-off as the limited capacity of the generator would have to synthesize all input sequences, instead of a subset of the input which is the case in IC-based training. Moreover, selective discrimination in IC helps the generator by making it harder for the discriminator to discriminate between real and fake sequences by imputing real sequences in its input. This prevents the discriminator to overpower the generator, thereby leading to a stable training process with an even learning ground. This phenomenon was not observed while training \mmgan without IC, in which without the selective discrimination in \D, it quickly overpowered the generator by easily detecting fakes and lead to overall poorer performance of the generator which was exacerbated by lack of selective loss computation in \G. In the final benchmark presented in Section VI-C of the manuscript, we train networks with implicit conditioning enabled.}

\section{Reconstruction Error in Other Planes}
\red{Due to the design of our proposed method in which 2D axial slices are used in training/testing, we ran another experiment in order to quantify whether any inconsistency in terms of reconstruction error (mean squared error) exists in other planes (sagittal and coronal). We calculated reconstruction errors for 5 test patients in LGG cohort for all three planes (axial, sagittal, and coronal). Then, we compared the error distributions from each plane using the Mann-Whitney U statistical test, which tests statistical significance between two unpaired sample distributions without the assumption that they are originally sampled from the Normal distribution. The null hypothesis for this test is chosen as follows: if an error value is randomly chosen from the first sample distribution, the value is equally likely to be either greater than or less than another random value chosen from the second sample distribution. In our case, we perform two tests; between axial and sagittal planes, and between axial and coronal planes. We maintain our choice of confidence threshold of 0.05 in this test as well. 
We report the results of the test in Suppl. Mat. Table S\ref{tab:pvalues}. We observe that out of the 10 observed p-values, all but one were significantly higher than our chosen confidence threshold of 0.05. Through these tests, we confirmed that the null hypothesis cannot be rejected due to high p-values in both the tests (axial versus sagittal and axial versus coronal), for all tested patients. \redtwo{Hence the test suggests that there may not be any significant differences or inconsistencies between reconstruction error distributions in different planes. }
}

% \clearpage
% Table generated by Excel2LaTeX from sheet 'Sheet1'
% \begin{landscape}
% Table generated by Excel2LaTeX from sheet 'Sheet1'
\begin{table*}[h]
\scriptsize
  \centering
  \caption{\red{Quantitative experiments testing three different imputation methods $z \in \{\text{average, noise, zeros}\}$, as well as two training strategies curriculum learning (CL) and random sampling (RS). The combination $z=$ zeros and curriculum learning based training performs the best in terms of MSE and PSNR, while $z=$ zeros and random sampling performs slightly better in SSIM. Bit strings 0001 to 1110 denote absence (0) or presence (1) of sequences in order \tone, \ttwo, \tonec, \ttwof respectively.}}
    \begin{tabular}{p{0.4em}>{\centering\arraybackslash}p{2.4em}>{\centering\arraybackslash}p{2.4em}>{\centering\arraybackslash}p{2.4em}>{\centering\arraybackslash}p{2.4em}>{\centering\arraybackslash}p{2.4em}>{\centering\arraybackslash}p{2.4em}>{\centering\arraybackslash}p{2.4em}>{\centering\arraybackslash}p{2.4em}>{\centering\arraybackslash}p{2.4em}>{\centering\arraybackslash}p{2.4em}>{\centering\arraybackslash}p{2.4em}>{\centering\arraybackslash}p{2.4em}>{\centering\arraybackslash}p{2.4em}>{\centering\arraybackslash}p{2.4em}>{\centering\arraybackslash}p{2.4em}>{\centering\arraybackslash}p{2.4em}>{\centering\arraybackslash}p{2.4em}}
    \toprule
    \boldmath{$z$} & \textbf{Strategy} & \textbf{Metrics} & \textbf{0001} & \textbf{0010} & \textbf{0011} & \textbf{0100} & \textbf{0101} & \textbf{0110} & \textbf{0111} & \textbf{1000} & \textbf{1001} & \textbf{1010} & \textbf{1011} & \textbf{1100} & \textbf{1101} & \textbf{1110} & \textbf{Mean}\\
    \midrule
    \multicolumn{1}{c}{\multirow{20}[3]{*}{\rotatebox[origin=c]{90}{\textbf{average}}}} & \multicolumn{1}{c}{\multirow{9}[3]{*}{\textbf{CL}}} & \multicolumn{1}{c}{\multirow{2}[2]{*}{\textbf{MSE}}} & 0.0102  $\pm$  0.0041 & 0.0072  $\pm$  0.0034 & 0.0043  $\pm$  0.0024 & 0.0136 $\pm$ 0.0049 & 0.0178 $\pm$ 0.0080 & 0.0088 $\pm$ 0.0040 & 0.0072 $\pm$ 0.0040 & 0.0135 $\pm$ 0.0064 & 0.0107 $\pm$ 0.0043 & 0.0104 $\pm$ 0.0039 & 0.0058 $\pm$ 0.0050 & 0.0148 $\pm$ 0.0058 & 0.0121 $\pm$ 0.0031 & 0.0125 $\pm$ 0.0037 & 0.0106 $\pm$ 0.0045 \\
    \cmidrule{3-18}
          &       & \multicolumn{1}{c}{\multirow{2}[2]{*}{\textbf{PSNR}}} & 24.3600 $\pm$ 3.1202 & 25.8186 $\pm$ 2.8329 & 26.5299 $\pm$ 2.2198 & 22.3548 $\pm$ 3.2600 & 21.5038 $\pm$ 2.6834 & 24.3666 $\pm$ 2.4599 & 25.3814 $\pm$ 2.2801 & 23.1930 $\pm$ 3.6652 & 23.8125 $\pm$ 2.1223 & 23.7930 $\pm$ 2.2222 & 25.5398 $\pm$ 2.7827 & 21.5824 $\pm$ 2.2552 & 22.5795 $\pm$ 1.2743 & 22.5779 $\pm$ 1.3326 & 23.8138 $\pm$ 2.4650 \\
    \cmidrule{3-18}
          &       & \multicolumn{1}{c}{\multirow{2}[2]{*}{\textbf{SSIM}}} & 0.8688 $\pm$ 0.0707 & 0.8890 $\pm$ 0.0495 & 0.9136 $\pm$ 0.0169 & 0.8642 $\pm$ 0.0655 & 0.8740 $\pm$ 0.0398 & 0.8982 $\pm$ 0.0174 & 0.9273 $\pm$ 0.0128 & 0.8853 $\pm$ 0.0507 & 0.9007 $\pm$ 0.0243 & 0.8748 $\pm$ 0.0186 & 0.9023 $\pm$ 0.0303 & 0.8844 $\pm$ 0.0190 & 0.9078 $\pm$ 0.0236 & 0.8676 $\pm$ 0.0198 & 0.8899 $\pm$ 0.0328 \\
\cmidrule{2-18}          & \multicolumn{1}{c}{\multirow{9}[3]{*}{\textbf{RS}}} & \multicolumn{1}{c}{\multirow{2}[2]{*}{\textbf{MSE}}} & 0.0091 $\pm$ 0.0036 & 0.0071 $\pm$ 0.0034 & 0.0047 $\pm$ 0.0028 & 0.0125 $\pm$ 0.0053 & 0.0153 $\pm$ 0.0069 & 0.0093 $\pm$ 0.0044 & 0.0082 $\pm$ 0.0044 & 0.0134 $\pm$ 0.0068 & 0.0113 $\pm$ 0.0047 & 0.0107 $\pm$ 0.0038 & 0.0061 $\pm$ 0.0052 & 0.0147 $\pm$ 0.0054 & 0.0129 $\pm$ 0.0038 & 0.0124 $\pm$ 0.0039 & 0.0105 $\pm$ 0.0046 \\
\cmidrule{3-18}
          &       & \multicolumn{1}{c}{\multirow{2}[2]{*}{\textbf{PSNR}}} & 24.7807 $\pm$ 3.0408 & 25.8981 $\pm$ 3.0860 & 26.4938 $\pm$ 2.3945 & 22.9278 $\pm$ 3.3844 & 22.2024 $\pm$ 2.7340 & 24.1413 $\pm$ 2.5214 & 24.8613 $\pm$ 2.4021 & 23.2776 $\pm$ 3.8887 & 23.7212 $\pm$ 2.3412 & 23.7417 $\pm$ 2.2673 & 25.5211 $\pm$ 2.8685 & 21.5878 $\pm$ 2.1707 & 22.4448 $\pm$ 1.2573 & 22.5763 $\pm$ 1.3752 & 23.8697 $\pm$ 2.5523 \\
          \cmidrule{3-18}
          &       & \multicolumn{1}{c}{\multirow{2}[2]{*}{\textbf{SSIM}}} & 0.8707 $\pm$ 0.0674 & 0.8892 $\pm$ 0.0497 & 0.9130 $\pm$ 0.0194 & 0.8709 $\pm$ 0.0609 & 0.8806 $\pm$ 0.0361 & 0.8969 $\pm$ 0.0193 & 0.9219 $\pm$ 0.0157 & 0.8835 $\pm$ 0.0544 & 0.8992 $\pm$ 0.0269 & 0.8751 $\pm$ 0.0184 & 0.9027 $\pm$ 0.0308 & 0.8844 $\pm$ 0.0186 & 0.9055 $\pm$ 0.0266 & 0.8691 $\pm$ 0.0209 & 0.8902 $\pm$ 0.0332 \\[0.3mm]
\cmidrule{1-18}    \multicolumn{1}{c}{\multirow{20}[3]{*}{\rotatebox[origin=c]{90}{\textbf{noise}}}} & \multicolumn{1}{c}{\multirow{9}[3]{*}{\textbf{CL}}} & \multicolumn{1}{c}{\multirow{2}[2]{*}{\textbf{MSE}}} & 0.0087 $\pm$ 0.0039 & 0.0078 $\pm$ 0.0033 & 0.0042 $\pm$ 0.0025 & 0.0108 $\pm$ 0.0049 & 0.0142 $\pm$ 0.0072 & 0.0092 $\pm$ 0.0049 & 0.0073 $\pm$ 0.0040 & 0.0124 $\pm$ 0.0065 & 0.0114 $\pm$ 0.0053 & 0.0110 $\pm$ 0.0036 & 0.0057 $\pm$ 0.0049 & 0.0153 $\pm$ 0.0063 & 0.0157 $\pm$ 0.0022 & 0.0118 $\pm$ 0.0051 & 0.0104 $\pm$ 0.0046 \\
\cmidrule{3-18}
          &       & \multicolumn{1}{c}{\multirow{2}[2]{*}{\textbf{PSNR}}} & 24.9639 $\pm$ 3.1463 & 25.4739 $\pm$ 2.8858 & 26.7597 $\pm$ 2.3988 & 23.9253 $\pm$ 3.5588 & 22.8694 $\pm$ 3.0571 & 24.4041 $\pm$ 2.4803 & 25.3652 $\pm$ 2.1933 & 23.4336 $\pm$ 3.7283 & 23.7107 $\pm$ 2.3808 & 23.5981 $\pm$ 2.2295 & 25.4479 $\pm$ 2.7345 & 21.7608 $\pm$ 2.4545 & 21.6432 $\pm$ 1.1993 & 22.9348 $\pm$ 1.6894 & 24.0208 $\pm$ 2.5812 \\
          \cmidrule{3-18}
          &       & \multicolumn{1}{c}{\multirow{2}[2]{*}{\textbf{SSIM}}} & 0.8747 $\pm$ 0.0647 & 0.8877 $\pm$ 0.0485 & 0.9133 $\pm$ 0.0192 & 0.8795 $\pm$ 0.0587 & 0.8858 $\pm$ 0.0346 & 0.8984 $\pm$ 0.0165 & 0.9238 $\pm$ 0.0119 & 0.8860 $\pm$ 0.0520 & 0.8997 $\pm$ 0.0266 & 0.8754 $\pm$ 0.0167 & 0.9020 $\pm$ 0.0289 & 0.8831 $\pm$ 0.0192 & 0.9008 $\pm$ 0.0262 & 0.8694 $\pm$ 0.0226 & 0.8914 $\pm$ 0.0319 \\[0.3mm]
\cmidrule{2-18}          & \multicolumn{1}{c}{\multirow{9}[3]{*}{\textbf{RS}}} & \multicolumn{1}{c}{\multirow{2}[2]{*}{\textbf{MSE}}} & 0.0094 $\pm$ 0.0034 & 0.0087 $\pm$ 0.0034 & 0.0039 $\pm$ 0.0019 & 0.0121 $\pm$ 0.0050 & 0.0141 $\pm$ 0.0061 & 0.0102 $\pm$ 0.0040 & 0.0057 $\pm$ 0.0036 & 0.0138 $\pm$ 0.0069 & 0.0116 $\pm$ 0.0040 & 0.0123 $\pm$ 0.0032 & 0.0059 $\pm$ 0.0057 & 0.0151 $\pm$ 0.0053 & 0.0158 $\pm$ 0.0014 & 0.0129 $\pm$ 0.0055 & 0.0108 $\pm$ 0.0042 \\
\cmidrule{3-18}
          &       & \multicolumn{1}{c}{\multirow{2}[2]{*}{\textbf{PSNR}}} & 24.7128 $\pm$ 3.1711 & 25.1588 $\pm$ 3.1052 & 26.7751 $\pm$ 2.2401 & 23.3274 $\pm$ 3.5124 & 22.7595 $\pm$ 2.6671 & 23.9286 $\pm$ 2.2215 & 26.2113 $\pm$ 1.8840 & 23.1377 $\pm$ 3.8143 & 23.5335 $\pm$ 1.9258 & 23.3185 $\pm$ 1.9840 & 25.5491 $\pm$ 2.8145 & 21.4729 $\pm$ 1.8973 & 21.7065 $\pm$ 0.8774 & 22.6359 $\pm$ 1.6579 & 23.8734 $\pm$ 2.4123 \\
          \cmidrule{3-18}
          &       & \multicolumn{1}{c}{\multirow{2}[2]{*}{\textbf{SSIM}}} & 0.8663 $\pm$ 0.0683 & 0.8838 $\pm$ 0.0535 & 0.9140 $\pm$ 0.0159 & 0.8740 $\pm$ 0.0600 & 0.8831 $\pm$ 0.0345 & 0.8942 $\pm$ 0.0144 & 0.9300 $\pm$ 0.0115 & 0.8855 $\pm$ 0.0515 & 0.8999 $\pm$ 0.0220 & 0.8742 $\pm$ 0.0178 & 0.9023 $\pm$ 0.0301 & 0.8832 $\pm$ 0.0167 & 0.9001 $\pm$ 0.0227 & 0.8691 $\pm$ 0.0227 & 0.8900 $\pm$ 0.0315 \\[0.3mm]
\cmidrule{1-18}    \multicolumn{1}{c}{\multirow{20}[3]{*}{\rotatebox[origin=c]{90}{\textbf{zeros}}}} & \multicolumn{1}{c}{\multirow{9}[3]{*}{\textbf{CL}}} & \multicolumn{1}{c}{\multirow{2}[2]{*}{\textbf{MSE}}} & 0.0092 $\pm$ 0.0037 & 0.0072 $\pm$ 0.0032 & 0.0040 $\pm$ 0.0022 & 0.0108 $\pm$ 0.0044 & 0.0129 $\pm$ 0.0054 & 0.0084 $\pm$ 0.0036 & 0.0061 $\pm$ 0.0035 & 0.0120 $\pm$ 0.0063 & 0.0109 $\pm$ 0.0040 & 0.0102 $\pm$ 0.0030 & 0.0057 $\pm$ 0.0046 & 0.0128 $\pm$ 0.0048 & 0.0120 $\pm$ 0.0025 & 0.0113 $\pm$ 0.0040 & \textbf{0.0095} $\pm$ \textbf{0.0039} \\
\cmidrule{3-18}
          &       &\multicolumn{1}{c}{\multirow{2}[2]{*}{\textbf{PSNR}}} & 24.7832 $\pm$ 3.3197 & 25.7012 $\pm$ 2.9797 & 26.9985 $\pm$ 2.4075 & 23.6527 $\pm$ 3.4296 & 22.9238 $\pm$ 2.6554 & 24.7581 $\pm$ 2.2131 & 25.9841 $\pm$ 2.1926 & 23.6018 $\pm$ 3.8153 & 23.8408 $\pm$ 2.1715 & 23.9202 $\pm$ 2.0746 & 25.6005 $\pm$ 2.7909 & 22.1330 $\pm$ 1.8389 & 22.4980 $\pm$ 1.1684 & 23.0852 $\pm$ 1.5142 & \textbf{24.2487} $\pm$ \textbf{2.4694} \\
          \cmidrule{3-18}
          &       & \multicolumn{1}{c}{\multirow{2}[2]{*}{\textbf{SSIM}}} & 0.8758 $\pm$ 0.0651 & 0.8925 $\pm$ 0.0486 & 0.9179 $\pm$ 0.0178 & 0.8811 $\pm$ 0.0560 & 0.8873 $\pm$ 0.0337 & 0.8984 $\pm$ 0.0144 & 0.9288 $\pm$ 0.0137 & 0.8908 $\pm$ 0.0509 & 0.9028 $\pm$ 0.0244 & 0.8792 $\pm$ 0.0178 & 0.9030 $\pm$ 0.0303 & 0.8885 $\pm$ 0.0164 & 0.9086 $\pm$ 0.0253 & 0.8692 $\pm$ 0.0224 & 0.8946 $\pm$ 0.0312 \\[0.3mm]
\cmidrule{2-18}          & \multicolumn{1}{c}{\multirow{9}[3]{*}{\textbf{RS}}} & \multicolumn{1}{c}{\multirow{2}[2]{*}{\textbf{MSE}}} & 0.0083 $\pm$ 0.0034 & 0.0078 $\pm$ 0.0025 & 0.0039 $\pm$ 0.0021 & 0.0114 $\pm$ 0.0047 & 0.0141 $\pm$ 0.0070 & 0.0094 $\pm$ 0.0046 & 0.0070 $\pm$ 0.0037 & 0.0115 $\pm$ 0.0053 & 0.0104 $\pm$ 0.0044 & 0.0102 $\pm$ 0.0029 & 0.0051 $\pm$ 0.0035 & 0.0142 $\pm$ 0.0052 & 0.0136 $\pm$ 0.0037 & 0.0122 $\pm$ 0.0043 & 0.0099 $\pm$ 0.0041 \\
\cmidrule{3-18}
          &       & \multicolumn{1}{c}{\multirow{2}[2]{*}{\textbf{PSNR}}} & 25.1025 $\pm$ 2.9154 & 25.5322 $\pm$ 2.7620 & 27.0696 $\pm$ 2.2400 & 23.4269 $\pm$ 3.2486 & 22.7479 $\pm$ 2.9075 & 24.2726 $\pm$ 2.3542 & 25.4209 $\pm$ 2.1961 & 23.6656 $\pm$ 3.5530 & 23.9555 $\pm$ 2.1310 & 23.9078 $\pm$ 1.9425 & 25.7566 $\pm$ 2.3658 & 21.7955 $\pm$ 2.1545 & 22.0829 $\pm$ 1.4224 & 22.6741 $\pm$ 1.5116 & 24.1008 $\pm$ 2.4075 \\
          \cmidrule{3-18}
          &       & \multicolumn{1}{c}{\multirow{2}[2]{*}{\textbf{SSIM}}} & 0.8814 $\pm$ 0.0608 & 0.8925 $\pm$ 0.0456 & 0.9193 $\pm$ 0.0173 & 0.8850 $\pm$ 0.0546 & 0.8879 $\pm$ 0.0352 & 0.9010 $\pm$ 0.0157 & 0.9264 $\pm$ 0.0125 & 0.8909 $\pm$ 0.0498 & 0.9042 $\pm$ 0.0256 & 0.8803 $\pm$ 0.0155 & 0.9050 $\pm$ 0.0281 & 0.8873 $\pm$ 0.0179 & 0.9063 $\pm$ 0.0258 & 0.8699 $\pm$ 0.0237 & \textbf{0.8955} $\pm$ \textbf{0.0306} \\[0.3mm]
    \bottomrule
    \end{tabular}%
  \label{tab:clnoclz}%
\end{table*}%
% \end{landscape}

% Table generated by Excel2LaTeX from sheet 'Sheet1'
\begin{table*}[h]
  \centering
  \caption{\red{Quantitative results for evaluating effectiveness of implicit conditioning (IC) based training strategy.}}
    \begin{tabular}{lcccccc}
    \toprule
          & \multicolumn{6}{c}{\textbf{Strategies}} \\
\cmidrule{2-7}    \textbf{Scenario} & \multicolumn{3}{c}{\textbf{No-IC}} & \multicolumn{3}{c}{\textbf{IC}} \\
\cmidrule{2-7} \tiny \tone \ttwo \tonec $T_{2f}$          & \textbf{MSE} & \textbf{PSNR} & \textbf{SSIM} & \textbf{MSE} & \textbf{PSNR} & \textbf{SSIM} \\
\cmidrule{2-7}    -\hspace{3mm}-\hspace{3mm}-\hspace{3mm}\checkmark   & 0.0120$\pm$0.0050 & 22.3268$\pm$3.0867 & 0.8282$\pm$0.0812 & 0.0092$\pm$0.0037 & 24.7832$\pm$3.3197 & 0.8758$\pm$0.0651 \\
    -\hspace{3.0mm}-\hspace{3.0mm}\checkmark\hspace{2mm}-   & 0.0086$\pm$0.0033 & 23.2450$\pm$2.3221 & 0.8456$\pm$0.0646 & 0.0072$\pm$0.0032 & 25.7012$\pm$2.9797 & 0.8925$\pm$0.0486 \\
    -\hspace{3.0mm}-\hspace{3.0mm}\checkmark\hspace{1.5mm}\checkmark   & 0.0036$\pm$0.0016 & 26.8574$\pm$1.9684 & 0.9092$\pm$0.0168 & 0.0040$\pm$0.0022 & 26.9985$\pm$2.4075 & 0.9179$\pm$0.0178 \\
    -\hspace{3.0mm}\checkmark\hspace{2.0mm}-\hspace{3.0mm}-   & 0.0089$\pm$0.0027 & 22.3974$\pm$2.3134 & 0.8294$\pm$0.0716 & 0.0108$\pm$0.0044 & 23.6527$\pm$3.4296 & 0.8811$\pm$0.0560 \\
    -\hspace{3.0mm}\checkmark\hspace{2.0mm}-\hspace{2.5mm}\checkmark   & 0.0152$\pm$0.0051 & 22.2603$\pm$2.4174 & 0.8749$\pm$0.0337 & 0.0129$\pm$0.0054 & 22.9238$\pm$2.6554 & 0.8873$\pm$0.0337 \\
    -\hspace{3.0mm}\checkmark\hspace{1.5mm}\checkmark\hspace{2.0mm}-    & 0.0076$\pm$0.0030 & 24.4887$\pm$2.2431 & 0.8912$\pm$0.0167 & 0.0084$\pm$0.0036 & 24.7581$\pm$2.2131 & 0.8984$\pm$0.0144 \\
    -\hspace{3.0mm}\checkmark\hspace{1.5mm}\checkmark\hspace{1.5mm}\checkmark   & 0.0054$\pm$0.0034 & 26.3621$\pm$1.8637 & 0.9304$\pm$0.0080 & 0.0061$\pm$0.0035 & 25.9841$\pm$2.1926 & 0.9288$\pm$0.0137 \\
     \checkmark\hspace{2.0mm}-\hspace{3.0mm}-\hspace{3.0mm}-   & 0.0116$\pm$0.0058 & 23.7413$\pm$3.6682 & 0.8858$\pm$0.0534 & 0.0120$\pm$0.0063 & 23.6018$\pm$3.8153 & 0.8908$\pm$0.0509 \\
     \checkmark\hspace{2.0mm}-\hspace{3.0mm}-\hspace{2.5mm}\checkmark   & 0.0119$\pm$0.0034 & 23.5427$\pm$1.8551 & 0.8999$\pm$0.0223 & 0.0109$\pm$0.0040 & 23.8408$\pm$2.1715 & 0.9028$\pm$0.0244 \\
    \checkmark\hspace{2.0mm}-\hspace{2.5mm}\checkmark\hspace{2.0mm}-   & 0.0090$\pm$0.0021 & 24.0322$\pm$1.6683 & 0.8698$\pm$0.0208 & 0.0102$\pm$0.0030 & 23.9202$\pm$2.0746 & 0.8792$\pm$0.0178 \\
     \checkmark\hspace{2.0mm}-\hspace{2.5mm}\checkmark\hspace{1.5mm}\checkmark   & 0.0050$\pm$0.0036 & 25.7482$\pm$2.2597 & 0.9013$\pm$0.0249 & 0.0057$\pm$0.0046 & 25.6005$\pm$2.7909 & 0.9030$\pm$0.0303 \\
    \checkmark\hspace{1.5mm}\checkmark\hspace{2.0mm}-\hspace{3.0mm}-   & 0.0133$\pm$0.0036 & 21.9508$\pm$1.5660 & 0.8855$\pm$0.0162 & 0.0128$\pm$0.0048 & 22.1330$\pm$1.8389 & 0.8885$\pm$0.0164 \\
     \checkmark\hspace{1.5mm}\checkmark\hspace{2.0mm}-\hspace{2.5mm}\checkmark   & 0.0188$\pm$0.0035 & 20.8677$\pm$1.3094 & 0.8957$\pm$0.0214 & 0.0120$\pm$0.0025 & 22.4980$\pm$1.1684 & 0.9086$\pm$0.0253 \\
    \checkmark\hspace{1.5mm}\checkmark\hspace{1.5mm}\checkmark\hspace{2mm}-   & 0.0094$\pm$0.0030 & 23.4557$\pm$1.3986 & 0.8671$\pm$0.0198 & 0.0113$\pm$0.0040 & 23.0852$\pm$1.5142 & 0.8692$\pm$0.0224 \\
    \midrule
    \textbf{mean$\pm$std} & 0.0100$\pm$0.0035 & 23.6626$\pm$2.1386 & 0.8796$\pm$0.0337 & \textbf{0.0095$\pm$0.0039} & \textbf{24.2487$\pm$2.4694} & \textbf{0.8946$\pm$0.0312} \\
    \bottomrule
    \end{tabular}%
  \label{tab:icnoic}%
\end{table*}%

% Please add the following required packages to your document preamble:
% \usepackage{booktabs}
% \usepackage{multirow}
\begin{table}[h]
\centering
\caption{\red{Results for Mann-Whitney U statistical test on 5 test patients from LGG cohort. $p$-values from two tests (axial vs coronal and axial vs sagittal) are reported.}}
\begin{tabular}{@{}lll@{}}
\toprule
\multicolumn{1}{c}{\multirow{2}{*}{\textbf{Patient Name}}} & \multicolumn{2}{c}{\textbf{p-values}}                                        \\ \cmidrule(l){2-3} 
\multicolumn{1}{c}{}                                       & \multicolumn{1}{c}{\textbf{Coronal}} & \multicolumn{1}{c}{\textbf{Sagittal}} \\ \midrule
Brats18\_2013\_9\_1                                        & 0.2814                               & 0.2714                                \\
Brats18\_TCIA10\_449\_1                                    & 0.4980                               & 0.4670                                \\
Brats18\_TCIA09\_451\_1                                    & 0.4848                               & 0.3592                                \\
Brats18\_TCIA12\_470\_1                                    & 0.4353                               & 0.0730                                \\
Brats18\_TCIA12\_466\_1                                    & 0.3061                               & 0.4213                                \\ \bottomrule
\end{tabular}
\label{tab:pvalues}%
\end{table}

% \bibliographystyle{IEEEtran}  
% \bibliography{library}  

% that's all folks